\documentclass[twocolumn,american]{revtex4-1}
\usepackage[T1]{fontenc}
\usepackage[latin9]{inputenc}
\usepackage{prettyref}
\usepackage{mathtools}
\usepackage{amsmath}
\usepackage{amssymb}
\usepackage{graphicx}

\makeatletter
\usepackage{filecontents}
\usepackage{xr}
\usepackage{hyperref}
\newrefformat{eq}{\hyperref[#1]{Eq.~(\ref{#1})}}
\newrefformat{cap}{\hyperref[#1]{Fig.~\ref{#1}}}
\newrefformat{fig}{\hyperref[#1]{Fig.~\ref{#1}}}
\newrefformat{tab}{\hyperref[#1]{Table ~\ref{#1}}}
\newrefformat{sec}{\hyperref[#1]{Section~\ref{#1}}}
\newrefformat{sub}{\hyperref[#1]{Section~\ref{#1}}}
\newrefformat{cha}{\hyperref[#1]{Chapter~\ref{#1}}}

\makeatother

\usepackage{babel}
\begin{document}
\global\long\def\o{\mathcal{O}}%

\title{Gell-Mann-Low criticality in neural networks} \author{Lorenzo Tiberi$^{1,2}$} \thanks{L. Tiberi and J. Stapmanns contributed equally to this work} \author{Jonas Stapmanns$^{1,2}$}  \thanks{L. Tiberi and J. Stapmanns contributed equally to this work} \author{Tobias K\"{u}hn$^{3}$} \author{Thomas Luu$^{4}$} \author{David Dahmen$^{1}$} \author{Moritz Helias$^{1,2}$} 
\affiliation{$^{1}$Institute of Neuroscience and Medicine (INM-6) and Institute for Advanced Simulation (IAS-6) and JARA-Institute Brain Structure-Function Relationships (INM-10),    Jülich Research Centre, Jülich, Germany } \affiliation{$^{2}$Institute for Theoretical Solid State Physics, RWTH Aachen University, 52074 Aachen, Germany}
\affiliation{$^{3}$Laboratoire de Physique de l'Ecole Normale Supérieure, ENS, Université PSL, CNRS, Sorbonne Université, Université de Paris, F-75005 Paris, France}
\affiliation{$^{4}$Institut für Kernphysik (IKP-3), Institute for Advanced Simulation (IAS-4) and Jülich Center for Hadron Physics, Jülich Research Centre, Jülich, Germany}
\begin{abstract}
Criticality is deeply related to optimal computational capacity. The
lack of a renormalized theory of critical brain dynamics, however,
so far limits insights into this form of biological information processing
to mean-field results. These methods neglect a key feature of critical
systems: the interaction between degrees of freedom across all length
scales, which allows for complex nonlinear computation. We present
a renormalized theory of a prototypical neural field theory, the stochastic
Wilson-Cowan equation. We compute the flow of couplings, which parameterize
interactions on increasing length scales. Despite similarities with
the Kardar-Parisi-Zhang model, the theory is of a Gell-Mann-Low type,
the archetypal form of a renormalizable quantum field theory. Here,
nonlinear couplings vanish, flowing towards the Gaussian fixed point,
but logarithmically slowly, thus remaining effective on most scales.
We show this critical structure of interactions to implement a desirable
trade-off between linearity, optimal for information storage, and
nonlinearity, required for computation.
\end{abstract}
\maketitle
Criticality and information processing are deeply related: Statistical
descriptions of the hardest-to-solve combinatorial optimization problems,
for example, are right at the edge of a phase transition \citep{Cheeseman91_331,Saitta11_book}.
Also brain activity shows criticality \citep{Beggs03_11167,Chialvo10_744},
which may optimize the network's computational properties. Among these,
long spatio-temporal correlations support information storage and
transmission, while rich collective and nonlinear dynamics allow degrees
of freedom to cooperatively perform complex signal transformations
\citep{Langton90_12}.

A multitude of critical processes could lie behind critical brain
dynamics. Thanks to the universality paradigm of statistical physics,
however, their macroscopic behavior is organized into few classes,
distinguished by only generic properties, like the system's dimension,
the nature of the degrees of freedom, as well as symmetries and conservation
laws \citep{Hohenberg77,Taeuber14}. Identifying the universality
classes implemented by neural networks and finding their distinctive
features is key to understanding if and how the brain exploits criticality
to perform computation.

Theoretical analysis of critical brain dynamics is, however, so far
restricted to mean-field methods. These partly explain why memory,
dynamic range and signal separation  are optimal at a critical point
\citep{Bertschinger04_1413,Bertschinger05,Kinouchi06_348}. Still,
a fundamental aspect of critical computation is inaccessible to these
methods: the nonlinear interaction between degrees of freedom across
all length scales. By approximating fluctuations as Gaussian, mean-field
theory can study only the linear response of individual modes to stimuli.
But a single, uncoupled mode can solve only simple computational tasks.
In fact, this approximation only holds if nonlinear interactions,
albeit necessary for computation, are irrelevant on macroscopic scales.
Thus, mean-field criticality is a special case, in fact the simplest
but also the most restricted kind of criticality the brain could possibly
implement. Uncovering different kinds of criticality requires more
sophisticated methods.

In this letter we analyze criticality in the stochastic Wilson-Cowan
rate model, a prototypical model of brain dynamics \citep{Wilson72_1,Wilson1973,Destexhe09_1}.
We use the non-equilibrium Wilsonian renormalization group  to go
beyond mean-field analysis \citep{Hohenberg77,Taeuber14}. This technique
tracks the flow of effective nonlinear couplings as one describes
the system on gradually increasing length scales. This exposes the
type of criticality featured by the system and its relevance for computation.
  The model is studied under a continuous stream of external inputs,
as typical for interconnected brain areas \citep{SchuezBraitenberg02,Destexhe03_739}.
While critical activity in networks driven by sparse inputs is well
described by branching processes \citep{Beggs03_11167,Haldeman05}
belonging to the universality class of mean-field directed percolation
\citep{Henkel08_book}, it is still unclear whether constantly driven
brain networks are able to support criticality at all \citep{Fosque21}.

We find the Wilson-Cowan model can indeed be critical in this regime.
The derived long-range description is very similar to that of the
Kardar-Parisi-Zhang model \citep{Kardar84}. But while the latter
features a strong coupling fixed point, we find a completely different
form of criticality: It is of the Gell-Mann-Low type, the archetypal
criticality underlying renormalizability of quantum field theories
\citep{GellMann1954}\citep[Sec. V]{Wilson75_773}. This type of
criticality occurs at the upper critical dimension, at which mean-field
theory looses validity. We determine this to be $d=2$, in agreement
with the planar organization of cortical networks. Nonlinear couplings
decrease only logarithmically slowly, thus remaining effective on
practically all length scales. This property implements a desirable
balance between a linear behavior and a nonlinear one; the model optimally
remembers signals presented in the past, due to its nearly linear
dynamics, and at the same time can perform nonlinear classification.

Focusing on the key ingredients of neural networks, that are nonlinear
dynamics of their constituents, noisy drive, and spatially localized
nonlinear coupling, we consider a neural field following the well-known
Wilson-Cowan equation
\begin{equation}
\tau\frac{dh}{dt}=-l\left(h\right)+w\ast f\left(h\right)+\sqrt{\tau}\,I\,,\label{eq:SDE}
\end{equation}
where $h\left(x,t\right)$ is a neural activity field on the space
$x\in\mathbb{R}^{d}$ that evolves in time $t$ on the characteristic
time-scale $\tau$. The function $l$ describes intrinsic local dynamics,
and $f$ is a nonlinear gain function. The connectivity kernel $w\left(x-x^{\prime}\right)$
weighs the input from the neural state at position $x^{\prime}$ to
that at position $x$, and $\ast$ is the spatial convolution. Following
a common approach \citep{Ermentrout98b}, the connectivity $w$ is
the sum of two Gaussians with widths $\sigma_{\pm}$ and amplitudes
$w_{\pm}$, with $\text{sign}\left(w_{\pm}\right)=\pm1$, representing
excitatory and inhibitory connections. External input from remote
brain areas driving the local activity is, for simplicity, modeled
as Gaussian white noise with statistics $\langle I\left(x,t\right)\rangle=\mu$
and $\langle I\left(x,t\right)I\left(x^{\prime},t^{\prime}\right)\rangle=D\delta\left(x-x^{\prime}\right)\,\delta\left(t-t^{\prime}\right)$.
A microscopic length $a$ characterizes the spatial resolution of
the model, thus \prettyref{eq:SDE} is defined on a square lattice
with $N^{d}$ sites and spacing $a$, eventually taking the limit
$N\to\infty$. Equivalently, momenta are restricted to $\left|k\right|<\Lambda\coloneqq\pi/a$.

\begin{figure*}
\begin{centering}
\includegraphics[width=1\textwidth]{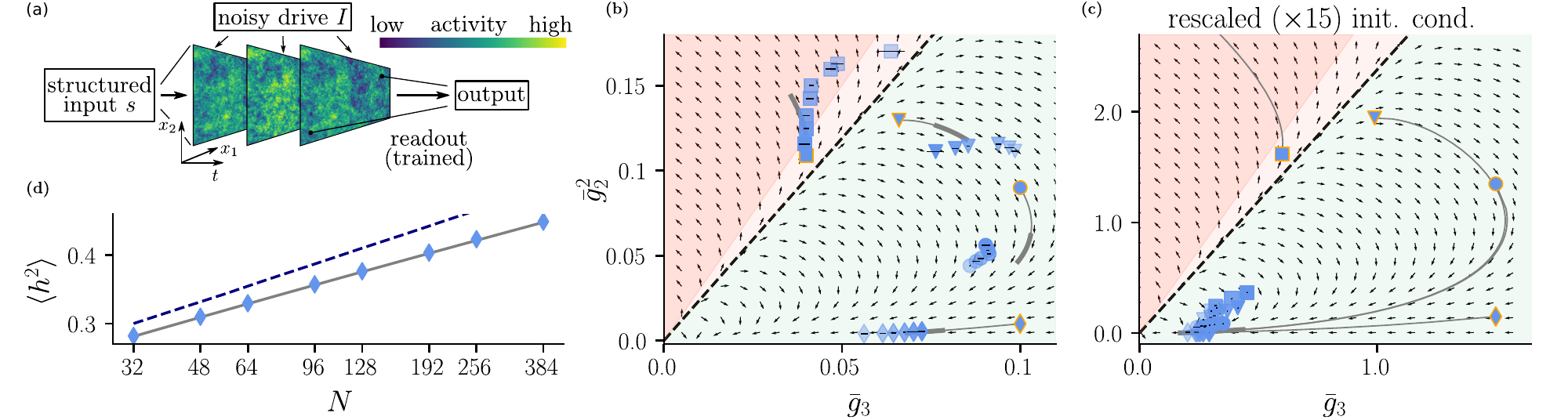}
\par\end{centering}
\caption{Phase diagram and flow of couplings. (a) Scheme of the model. (b)
and (c): Direction of flow (arrows) in the $\bar{g}_{3}-\bar{g}_{2}^{2}$
plane, as defined by \prettyref{eq:flow_gamma} and \prettyref{eq:flow_eta}.
The transition line (black, dashed) separates the plane in a converging
(green) and diverging region (red). Region of non-physical bi-stable
regime in darker red (see Supplemental Material). Blue markers represent
the measured flow of couplings at $\ell=\frac{N}{2}$, with $N\in\left\{ 32,48,64,96,128,192\right\} $
(dark to light shading; squares in (b) also include $N=256,384$)
for different initial conditions (yellow edge). Gray solid lines (thicker
within the range of simulated $\ell$) represent the corresponding
predicted flow. (d):\textbf{ }Measured variance (blue diamonds) as
a function of $N$ for $\bar{g}_{3}=0.1$ and $\bar{g}_{2}^{2}=0.01$
($\bar{g}_{2}^{2}\ll\bar{g}_{3}$ regime), compared against the linear
case (dashed, dark blue line) and prediction (gray, solid line). \label{fig:Phase-diagram}}
\end{figure*}

The computational properties of the model are later tested in a reservoir
computing setting \citep{Lukovsevivcius09}: A linear readout is trained
to extract a desired input-output mapping from the neural activity
(\prettyref{fig:Phase-diagram}(a)). Nonlinear interactions are fundamental
to achieve complex mappings. We thus want to go beyond mean-field
methods and instead track the relevance of nonlinear interactions
on gradually increasing length scales.

We make explicit all nonlinear terms in \prettyref{eq:SDE} by Taylor
expanding $f\left(x\right)=\sum_{n}f_{n}x^{n}$ and likewise $l$.
Also, the momentum dependence of the Fourier transformed coupling
kernel $\hat{w}\left(k\right)=\sum_{\pm}w_{\pm}\left(1-\frac{1}{2}\sigma_{\pm}^{2}k^{2}+\mathit{\o}\left(k^{4}\right)\right)$
is kept up to second order, enough to expose those terms that characterize
the system on a mesoscopic length-scale. We thus obtain, in the spatial
domain,
\begin{equation}
\tau\frac{dh}{dt}=\sum_{n=1}^{\infty}\left(-m_{n}+g_{n}\Delta+\o\left(\Delta^{2}\right)\right)\,h^{n}+I\,,\label{eq:SDE_taylor}
\end{equation}
where $\Delta$ is the Laplace operator and $\mathit{\o}\left(\Delta^{2}\right)$
denotes all terms proportional to spatial differential operators of
order $\geq4$. Couplings are conveniently renamed $m_{n}\coloneqq l_{n}-f_{n}\sum_{\pm}w_{\pm}$,
characterizing the local dynamics, and $g_{n}\coloneqq\frac{1}{2}f_{n}\sum_{\pm}\sigma_{\pm}^{2}w_{\pm}$,
quantifying the interaction across space points. Notice $\mu$, $l_{0}$
and $f_{0}$, corresponding to a constant input causing $\langle h\rangle\neq0$,
are without loss of generality set to zero (see Supplemental Material).

Temporarily neglecting nonlinearities ($g_{n}$, $m_{n}=0$ $\forall n>1$),
the mass term $m_{1}$ plays the role of a lower momentum cut-off
\citep{Wilson75_773}. This defines the system's spatial ($\propto m_{1}^{-\frac{1}{2}}$)
and temporal ($\propto m_{1}^{-1}$) correlation lengths, both diverging
as $m_{1}\to0$. Thus $m_{1}=0$ identifies a critical point of the
linear model. To include the effect of nonlinearities close to this
point, standard expansion methods fail: perturbative corrections diverge
due to statistical fluctuations interacting on an infinite range of
length scales \citep{Wilson75_773}.

To tackle this issue, the renormalization group (RG) \citep{Wilson75_773,Hohenberg77,Taeuber14}
performs the integration of fluctuations gradually over momentum scales
$\Lambda/\ell<\left|k\right|<\Lambda$; progressively increasing the
flow parameter $\ell\in[1,\infty)$, we obtain a series of effective
field theories, each describing only degrees of freedom on larger
length scales, with $k<\Lambda/\ell$. Such theories are defined on
the rescaled quantities $k_{\ell}\coloneqq\ell k$, $t_{\ell}\coloneqq\ell^{-z}t$,
$\hat{h}_{\ell}\left(k_{\ell},t_{\ell}\right)\coloneqq\ell^{-\zeta}\hat{h}\left(k,t\right)$
and $\hat{I}_{\ell}\left(k_{\ell},t_{\ell}\right)\coloneqq\ell^{\chi}\hat{I}\left(k,t\right)$.
Rescaling makes all effective field theories look formally equivalent
to \prettyref{eq:SDE_taylor}, differing only by the values of the
couplings $m_{n}\left(\ell\right)$ and $g_{n}\left(\ell\right)$,
which become $\ell$-dependent. This dependence accounts for two effects:
The interaction with the integrated out degrees of freedom and the
rescaling. The couplings' flow with $\ell$ therefore characterizes
nonlinear interactions on different length scales and, thus, the type
of critical behavior featured by the system. For example, the flow
running into a fixed point is a characteristic of critical systems
and determines their typical scale invariance.

We begin with analyzing the coupling's flow due to rescaling alone.
This corresponds to standard dimensional analysis and the mean-field
approach, neglecting the contribution of fluctuations. As typical
brain networks have a planar organization, we are ultimately interested
in the case $d=2$. We choose $z=2,$ $\zeta=\frac{d+2}{2},$ and
$\chi=\frac{d-2}{2}$ so that $g_{1}$, $\tau$, and the input variance
$D$ are at a fixed point (i.e. do not rescale). With this choice,
for $d\geq2$ all couplings not appearing explicitly in \prettyref{eq:SDE_taylor}
rapidly flow to $0$ with some negative power of $\ell$. These couplings
are termed \emph{irrelevant} and can be neglected as the effective
theory's reference scale $\ell$ is increased. The couplings $m_{n}\left(\ell\right)=\ell^{2-\left(n-1\right)\frac{d-2}{2}}m_{n}\left(1\right)$
diverge at $d=2$ as $\sim\ell^{2}$. They are termed \emph{relevant},
meaning they must be fine-tuned to $0$ to be at a fixed point. This
fine-tuning here implies balance of inhibitory and excitatory inputs
(see Supplemental Material), often observed in brain networks as a
necessary condition for criticality \citep{Shew09}. The couplings
$g_{n}\left(\ell\right)=\ell^{-\left(n-1\right)\frac{d-2}{2}}g_{n}\left(1\right)$
vanish for $d>2$, $\forall n\geq2$. For these spatial dimensions,
mean-field theory is usually accurate: all nonlinear terms in \prettyref{eq:SDE_taylor}
are negligible at large scales, thus fluctuations have almost no interactions
and can be neglected. Conversely, dimensional analysis predicts $d=2$
as the upper critical dimension at which the $g_{n}$ do not scale
and are thus termed \emph{marginal}: their flow is driven by fluctuations
alone and thus must be investigated with more sophisticated methods,
like the RG.

The mean-field analysis above allows us to determine the form of the
effective theory describing the critical system at a mesoscopic scale,
where irrelevant couplings are negligible
\begin{equation}
\tau\frac{dh}{dt}=\Delta\left(g_{1}h+g_{2}h^{2}+g_{3}h^{3}\right)+I\,.\label{eq:SDE_mesoscopic}
\end{equation}
Notice that, at such scales, the field $h$ describes neural populations
exchanging activity with nearest neighbors via a diffusive process,
as expressed by the Laplace operator $\Delta$ \citep{diSanto18_1356}.
Among the marginal couplings $g_{n}$, we keep only the first $n\leq n_{0}=3$.
We can assume neural activity to mainly explore a limited range of
the gain function \citep{Ostojic11_e1001056,Roxin11_16217}, which
can therefore be locally approximated with a polynomial. We choose
$n_{0}=3$ to keep a minimal approach ($n_{0}=2$ cannot be chosen,
as \prettyref{eq:SDE_mesoscopic} would be unstable). Equation \eqref{eq:SDE_mesoscopic}
was proposed as an alternative to the Kardar-Parisi-Zhang (KPZ) model
\citep{Pavlik94_553,Antonov95_485}, both describing the dynamic growth
of interfaces. The original KPZ model \citep{Kardar84} defines the
KPZ universality class, where the interaction flows into a strong-coupling
fixed point. Despite the similarities, we show \prettyref{eq:SDE_mesoscopic}
to exhibit a radically different type of critical behavior.

We start from \prettyref{eq:SDE_mesoscopic} to compute the fluctuation-driven
part of the couplings' flow. This is conveniently done by mapping
\prettyref{eq:SDE_mesoscopic} to a field theory by means of the Martin-Siggia-Rose-de~Dominicis-Janssen
formalism \citep{Martin73,dedominicis1976_247,Hertz16_033001,Helias20_970}.
Flow equations are then computed to one-loop order with Feynman diagram
techniques, within the framework of the infinitesimal momentum shell
Wilsonian RG for non-equilibrium systems \citep{Hohenberg77} (details
in Supplemental Material). Notice only $g_{2}$ and $g_{3}$ are meaningful
parameters, as one can rewrite \prettyref{eq:SDE_mesoscopic} in dimensionless
form by setting $\tau,\,g_{1},\,D$ and $a$ to unity and renaming
$\bar{g}_{n}\coloneqq\left(D/g_{1}\right)^{\frac{n-1}{2}}\frac{g_{n}}{g_{1}}$.
We use this convention for the remainder of this letter. Defining
$s\coloneqq\left(2\pi\right)^{-1}\ln\left(\ell\right)$, the differential
flow equations take the form
\begin{gather}
\frac{1}{g_{1}}\frac{dg_{1}}{ds}=\frac{3}{2}\bar{g}_{3}-\bar{g}_{2}^{2},\label{eq:flow_nu}\\
\frac{d\bar{g}_{2}^{2}}{ds}=-\frac{27}{2}\bar{g}_{3}\bar{g}_{2}^{2}+7\bar{g}_{2}^{4},\label{eq:flow_gamma}\\
\frac{d\bar{g}_{3}}{ds}=-\frac{15}{2}\bar{g}_{3}^{2}+14\bar{g}_{3}\bar{g}_{2}^{2}-4\bar{g}_{2}^{4}\,,\label{eq:flow_eta}
\end{gather}
showing that $\bar{g}_{2}$ and $\bar{g}_{3}$ alone drive the flow.
The Laplace operator in front of the nonlinear terms in \prettyref{eq:SDE_mesoscopic}
protects $D$, $\tau$ and $m_{n}$ from fluctuation corrections (see
Supplemental Material), so their flow is completely determined by
the mean-field analysis above. \prettyref{fig:Phase-diagram}(b)\textbf{
}shows the flow vector field in the $\bar{g}_{3}-\bar{g}_{2}^{2}$
plane. A line $\frac{\bar{g}_{2}^{2}}{\bar{g}_{3}}=\frac{7+\sqrt{145}}{8}$
determines a transition between a diverging and converging behavior.

Below the transition line, the couplings vanish, flowing into the
Gaussian fixed point $(\bar{g}_{2}=\bar{g}_{3}=0$). Differently than
in a mean-field scenario, however, the flow is logarithmically slow
in $\ell$, a characteristic of marginal couplings at the upper critical
dimension. This means interactions are effectively present on a wide
range of length scales. Finite-size systems typically do not even
reach the extent beyond which interactions become truly negligible.
This is known as Gell-Mann-Low criticality, the archetypal behavior
underlying renormalizability of quantum field theories, such as Quantum
Electrodynamics (QED) \citep{GellMann1954}\citep[Sec. V]{Wilson75_773}.
Here one observes nonlinearities that shape non-trivial electro-magnetic
interactions on a wide range of scales. A difference to a prototypical
Gell-Mann-Low theory, however, is that the flow is driven by the
pair of marginal couplings $(\bar{g}_{2},\bar{g}_{3})$, rather than
by a single coupling, which for QED is the charge. At large scales,
the power law scaling exponents $z,\,\zeta$, and $\chi$ maintain
their mean-field values, since interactions eventually vanish. However,
logarithmic corrections must be included due to the slowness of the
flow. One example of this effect can be seen in the scaling of the
variance $\big\langle h^{2}\big\rangle$ as a function of $N$ shown
in \prettyref{fig:Phase-diagram}(d). These logarithmic corrections
may also be mistaken as different power law exponents, which depend
on the system's dynamical state, rather than being universal. Thus
they can lead to small, state-dependent shifts in measured scaling
exponents with respect to their mean-field value (see Supplemental
Material).

As previously noted by \citep{Antonov95_485}, the infinite number
of marginal couplings $g_{n}$ in principle allows for the existence
of an infinite number of fixed points. Determining analytically whether
these are attractive, though, has never been done due to the technical
difficulty of the task. In neglecting $g_{n}$, $\forall n>3$, we
are implicitly assuming the Gaussian fixed point is stable, with a
sufficiently large basin of attraction to attract the flow from any
initial $g_{n}$. We therefore test the validity of the flow equations
numerically, by integrating \prettyref{eq:SDE_mesoscopic} with the
Euler-Maruyama algorithm \citep{Kloeden92}(additional details in
Supplemental Material). Measurements are restricted to $\bar{g}_{2}^{2}<3\bar{g}_{3}$
(\prettyref{fig:Phase-diagram}(b)), due to the occurrence of an unphysical
bi-stable regime above such boundary (Supplemental Material). 

We simulate systems of increasing size $N$, which limits the extent
of correlations. By measuring correlation functions of the neural
field $h$, we extract the value of the flown couplings at different
length scales $\ell=\frac{N}{2}$ and initial conditions, shown in
\prettyref{fig:Phase-diagram}(b). Below the transition line, we observe
good qualitative agreement between the measured and predicted flow.
Quantitative departures from predictions are expected, given the approximation
made to one-loop order in fluctuations and to third order in the expansion
of $f$. This approximation is good enough, given our goal to qualitatively
confirm the running of the flow towards the Gaussian fixed point.
Higher orders could be easily included, if needed, at the relatively
low cost of computing more Feynman diagrams.

Above the transition line instead, our approximation leads to a divergence
of the flow. This could signal the presence of a strong coupling fixed
point, into which the flow eventually runs. This occurs, for example,
in the KPZ model for $d\geq3$, where an analogous transition point
exists \citep{Fogedby09}. Differently, however, our measurements
in \prettyref{fig:Phase-diagram}(b) suggest the flow still heads
towards the Gaussian fixed point, making a u-turn similar to when
starting close and below the transition line. This is confirmed in
\prettyref{fig:Phase-diagram}(c), showing the flow for larger initial
values of the couplings. The flow is then subject to a stronger drive
in its initial transient, thus spanning a longer trajectory within
the same range of $\ell$. These measurements thus suggest that, regardless
of the initial condition, after an initial transient the flow always
runs into the $\bar{g}_{3}\gg\bar{g}_{2}^{2}$ region, heading towards
the Gaussian fixed point. Luckily, this region is where the flow equations
in the given approximation yield reliable quantitative predictions.
This is exemplified by the measured neural field variance $\big\langle h^{2}\big\rangle$
as a function of $N$ being well predicted by theory (\prettyref{fig:Phase-diagram}(d)).
In the limit $N\to\infty$, we predict $\big\langle h^{2}\big\rangle\sim\left(4\pi g_{1}\left(\ell\right)\right)^{-1}\ln\left(\frac{N}{2}\right)$,
with $\ell\overset{!}{=}\frac{N}{2}$ (see Supplemental Material).
The $\ell$-dependence of $g_{1}$ shows a logarithmic correction
with respect to the linear case, in which $g_{1}\left(\ell\right)=1$
$\forall\ell$. 
\begin{figure*}
\begin{centering}
\includegraphics[width=1\textwidth]{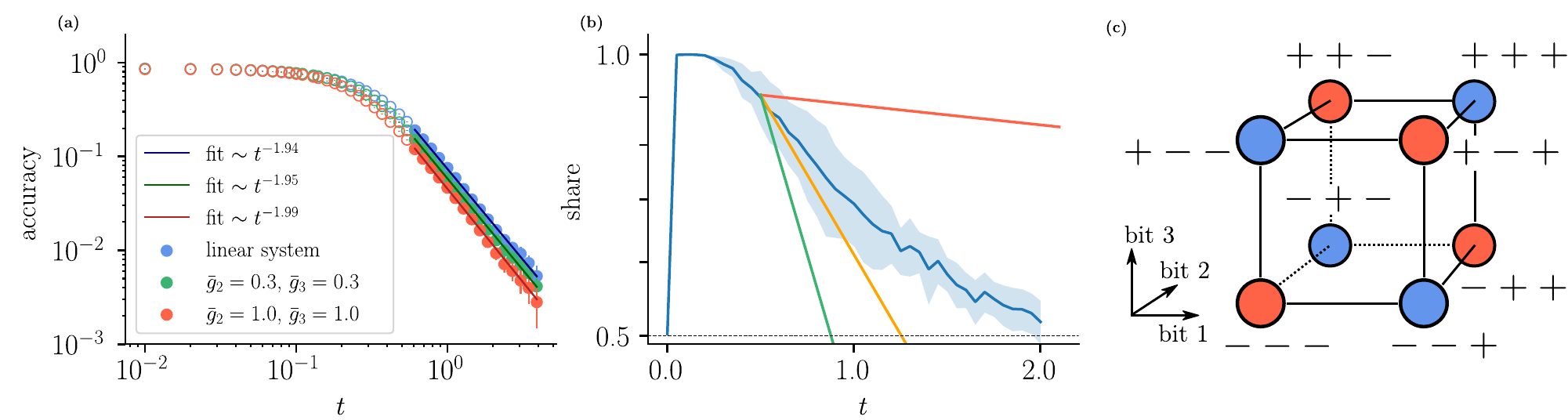}
\par\end{centering}
\caption{Examples of computation. (a) Memory: reconstruction accuracy over
time of a Gaussian input for different strengths of the nonlinearties
(legend). Values range from $0$ (complete forgetting) to $1$ (perfect
reconstruction).  (b)\textbf{ }Classification: Share of correctly
classified $3$-bit strings (blue line) and standard deviation across
repeated trainings (blue area). Other curves decay with the time-scale
of the slowest (red), middle (orange) and fastest (green) modes available
at readout. (c) Spatial position of the $3$-bit strings, colored
by their parity. \label{fig:Examples-of-computation.}}
\end{figure*}

We have so far demonstrated that the system showcases criticality
of the Gell-Mann-Low kind. The question remains on why, among types
of criticality, this one would be especially beneficial for computation.
Criticality optimizes desirable computational properties; still, these
may strive for a balance between a linear behavior, optimal for storage
and transmission of information, and a nonlinear one, necessary for
computation. The Gell-Mann-Low criticality implements such a balance
by sitting in-between a mean-field and strong coupling fixed point
scenario. We exemplify this by training the system \prettyref{eq:SDE_mesoscopic}
to solve example tasks by what is known as reservoir computing (\prettyref{fig:Phase-diagram}(a))
\citep{Lukovsevivcius09}: A structured input $s$ is fed at some
time $t_{\mathit{in}}$ in the form of a perturbation $h\left(x,t_{\mathrm{in}}\right)+s\left(x\right)$.
At a later time $t_{\mathrm{out}}$, a linear readout $\sum_{x}W\left(x\right)h\left(x,t_{\mathrm{out}}\right)$
is taken and the parameters $W\left(x\right)$ are trained with gradient
descent (task-specific details in Supplemental Material).

We first focus on memory; for concreteness, consider the Fischer
memory curve \citep{Ganguli2008_18970}. In the linear case, it is
expected to be optimal, and to decay with time as $t^{-2}$. Being
a global feature, involving the system as a whole on long time scales,
it benefits from the system's closeness to linear dynamics. Indeed,
since nonlinearities vanish on a global length scale, as opposed to
a strong coupling fixed point scenario, they do not worsen the power
law decay found in the optimal linear case; rather they cause only
small logarithmic corrections. Inspired by these theoretical grounds,
we train $N^{2}$ linear readouts to reconstruct a Gaussian-shaped
input at some time $t$ after injection, recording the reconstruction
accuracy (\prettyref{fig:Examples-of-computation.}(a)). As expected,
increasing the strength of nonlinearities, the accuracy's power law
exponent does not deviate appreciably from the linear case: performance
is only worsened by a constant shift in double logarithmic scale.

In contrast to a strict mean-field scenario, nonlinearities are, however,
still relevant on most length scales other than the very macroscopic
ones, thus allowing the system's degrees of freedom to interact and
to collectively perform computation. We exemplify this by training
two linear readouts to correctly classify the parity of $3$-bit strings.
The task is not linearly separable: no plane can correctly separate
the strings in the two categories (\prettyref{fig:Examples-of-computation.}(c)).
Nonlinear dynamics are necessary to expand the input's dimensionality,
thus making linear separation possible. The input string's $n$-th
bit is encoded by the sign of a perturbation of the Fourier mode of
momentum $k_{n}=\left(\pi-n\,4\pi/N\right)\hat{e}_{1}$, right below
the high momentum cut-off $\pi$. The linear readout is taken on a
low-pass filtered $h$, restricted to momenta $<\left|k_{3}\right|$.
Thus nonlinear dynamics are further necessary to transfer information
from the input to the readout modes. \prettyref{fig:Examples-of-computation.}(b)
indeed shows successful training. Performance decays on a time scale
between the characteristic scales of the slowest and fastest modes,
suggesting their collective interaction to be employed to perform
the task. The modes' response shown in the Supplemental Material further
imply this, see \prettyref{fig:k_modes}.

In conclusion, we furnish the tools, a renormalized theory of neural
network dynamics, to uncover the structure of nonlinear interactions
across scales, so far inaccessible by mean-field methods. This framework
opens the door to explore many other forms of criticality, beyond
mean-field, that neural networks might implement, and to quantitatively
address nonlinear signal transformations at a dynamical critical point.
Applying the methods to the stochastic Wilson-Cowan model, we find
a new form of criticality, which is robust to the biologically plausible
external drive. At the relevant dimension $d=2$, critical behavior
ceases to be mean-field, being instead of the Gell-Mann-Low kind.
Most notably, nonlinear couplings, fundamental for computation, have
a logarithmically slow, rather than power-law vanishing flow. We argue
that this supports a computationally optimal balance between linear
and nonlinear dynamics. The RG methodology and the concept of Gell-Mann-Low
criticality are generally applicable to any stochastic dynamical system.
We thus foresee applications to computation-oriented, possibly more
complex models, both biological and artificial. For example, Gell-Mann-Low
criticality could naturally account for the variability in critical
exponents observed in the context of neural avalanches \citep{Fontenele19_208101,Carvalho21}.
The link between neural field theory, the KPZ model central to non-equilibrium
statistical physics, and quantum field theory presents a stepping
stone to transfer expertise from these fields, where RG methods are
widely used.

\begin{acknowledgments}
We are grateful for various helpful discussions with Carsten Honerkamp.
This project has received funding from the European Union\textquoteright s
Horizon 2020 Framework Programme for Research and Innovation under
Specific Grant Agreement No. 945539 (Human Brain Project SGA3); the
Helmholtz Association: Young Investigator's grant VH-NG-1028; the
Jülich-Aachen Research Alliance Center for Simulation and Data Science
(JARA-CSD) School for Simulation and Data Science (SSD); L.T received
funding as Vernetzungsdoktorand: \textquotedbl Collective behavior
in stochastic neuronal dynamics\textquotedbl .
\end{acknowledgments}

\bibliographystyle{apsrev_brain}
\bibliography{brain}

\makeatletter\@input{xsupplement.tex}\makeatother
\end{document}

\global\long\def\th{\tilde{h}}%
\global\long\def\tH{\tilde{H}}%
\global\long\def\T{\mathrm{T}}%

\global\long\def\tj{\tilde{j}}%
\begin{appendices}
\renewcommand{\thesection}{\Roman{section}}
\renewcommand{\thefigure}{S\arabic{figure}}
\renewcommand{\thetable}{S\arabic{table}}
\renewcommand{\theequation}{S\arabic{equation}}
\title{Supplemental material: Gell-Mann-Low criticality in neural networks} \author{Lorenzo Tiberi$^{1,2}$} \thanks{L. Tiberi and J. Stapmanns contributed equally to this work} \author{Jonas Stapmanns$^{1,2}$}  \thanks{L. Tiberi and J. Stapmanns contributed equally to this work} \author{Tobias K\"{u}hn$^{3}$} \author{Thomas Luu$^{4}$} \author{David Dahmen$^{1}$} \author{Moritz Helias$^{1,2}$} 
\affiliation{$^{1}$Institute of Neuroscience and Medicine (INM-6) and Institute for Advanced Simulation (IAS-6) and JARA-Institute Brain Structure-Function Relationships (INM-10),    Jülich Research Centre, Jülich, Germany } \affiliation{$^{2}$Institute for Theoretical Solid State Physics, RWTH Aachen University, 52074 Aachen, Germany}
\affiliation{$^{3}$Laboratoire de Physique de l'Ecole Normale Supérieure, ENS, Université PSL, CNRS, Sorbonne Université, Université de Paris, F-75005 Paris, France}
\affiliation{$^{4}$Institut für Kernphysik (IKP-3), Institute for Advanced Simulation (IAS-4) and Jülich Center for Hadron Physics, Jülich Research Centre, Jülich, Germany}
\let\clearpage\relax
\maketitle

\section{Mapping into a field theory}

Stochastic differential equations, the likes of \prettyref{eq:SDE},
can be equivalently represented in terms of a field theory by means
of the Martin-Siggia-Rose-de~Dominicis-Janssen (MSRDJ) formalism
\citep{Martin73,dedominicis1976_247,Hertz16_033001,Helias20_970}.
Here we give the field theory's action that can be derived with this
mapping, as well as its associated Feynman rules \citep{ZinnJustin96}.
The latter are used to compute the flow equations, as well as the
analytical predictions that are compared against numerical simulations.

\paragraph{Field theory}

The statistics of the neural field $h$ are described in terms of
a moment generating functional
\begin{equation}
Z\left[j,\tj\right]=\int\mathcal{D}h\mathcal{D}\th\exp\left(S\left(h,\th\right)+j^{\T}h+\tj^{\T}\th\right)\,,\label{eq:moment_generating_functional}
\end{equation}
defined by a path integral over the field $h$ and an auxiliary field
$\th$, with the action 
\begin{equation}
S\left(h,\th\right)=\th^{\T}\left(\frac{d}{dt}h+l\left(h\right)-w\ast f\left(h\right)\right)+W_{I}\left(-\th\right)\,,\label{eq:full_action}
\end{equation}
where we used the notation $f^{\T}g\coloneqq\int d^{d}x\,dt\,f\left(x,t\right)g\left(x,t\right)$
and we have set $\tau=1$ in \prettyref{eq:SDE}, corresponding to
a trivial rescaling of time. Here, $W_{I}$ is the cumulant generating
functional of the random input $I$. In our case
\begin{equation}
W_{I}\left(-\th\right)=-\mu^{\T}\th+\frac{D}{2}\th^{\T}\th\label{eq:noise_cumulant_generating_functional}
\end{equation}
with $\mu\left(r,t\right)\coloneqq\mu=\mathit{const}$. Moments of
$h$ and $\th$ correspond to functional derivatives of $Z$ with
respect to $n$ source fields $j$ and $m$ fields $\tj$. Specifically
\begin{align}
\frac{\partial Z}{\partial j\left(x_{1},t_{1}\right)\ldots\partial j\left(x_{n},t_{n}\right)\partial\tj\left(x'_{1},t'_{1}\right)\ldots\partial\tj\left(x'_{m},t'_{m}\right)}\Bigg|_{j,\tj=0} & =\frac{\partial\langle h\left(x_{1},t_{1}\right)\ldots h\left(x_{n},t_{n}\right)\rangle}{\partial\tj\left(x'_{1},t'_{1}\right)\ldots\partial\tj\left(x'_{m},t'_{m}\right)}\Bigg|_{\tj=0}\nonumber \\
 & =\langle h\left(x_{1},t_{1}\right)\ldots h\left(x_{n},t_{n}\right)\th(x'_{1},t'_{1})\ldots\th(x'_{m},t'_{m})\rangle\,.\label{eq:derivatives_moments}
\end{align}
Setting the number of $\tj$-derivatives to $m=0$, one can thus compute
moments of the neural field $h$. Additionally, differentiating $m$
times with respect to $\tj$ allows one to compute the $m$-th order
response function of such moments to a small perturbation in the system's
input. This can be seen from the fact that, in \prettyref{eq:moment_generating_functional},
both $\mu$ and $\tj$ couple linearly with $\th$. Thus $-\tj$ represents
an additive perturbation to the average input $\mu$. Notice that
\prettyref{eq:derivatives_moments} relates derivatives in $\tj$
to moments of $\th$, thus the statistics of the auxiliary field $\th$
has the physical interpretation of describing the system's response
to stimuli. For this reason $\th$ is usually called response field.
By construction instead, all moments of the field $\th$ alone ($n=0$
derivatives by $j$) vanish \citep[p. 38]{Coolen00_arxiv_II}.

\paragraph{Feynman rules}

The Feynman rules for the field theory \prettyref{eq:full_action}
are given in terms of propagators and interaction vertices in momentum
and frequency domain. We denote by $H$ and $\tilde{H}$ the Fourier
transformed fields $h$ and $\tilde{h}$ in both frequency and momentum
space. Propagators are given by inverting the second order functional
derivative of the action
\[
\left(\begin{array}{cc}
\Delta_{HH}\left(k,\omega\right) & \Delta_{H\tilde{H}}\left(k,\omega\right)\\
\Delta_{H\tilde{H}}\left(-k,-\omega\right) & \Delta_{\tilde{H}\tilde{H}}\left(k,\omega\right)
\end{array}\right)\left(\begin{array}{cc}
\frac{\partial^{2}S}{\partial H\left(k,\omega\right)\partial H\left(k',\omega'\right)} & \frac{\partial^{2}S}{\partial H\left(k,\omega\right)\partial\tilde{H}\left(k',\omega'\right)}\\
\frac{\partial^{2}S}{\partial\tilde{H}\left(k,\omega\right)\partial H\left(k',\omega'\right)} & \frac{\partial^{2}S}{\partial\tilde{H}\left(k,\omega\right)\partial\tilde{H}\left(k',\omega'\right)}
\end{array}\right)_{H,\tH=0}=-\left(2\pi\right)^{d+1}\delta\left(k+k'\right)\delta\left(\omega+\omega'\right)\:,
\]
which gives explicitly

\begin{fmffile}{tpo_propagator}
\fmfset{thin}{0.75pt}
\fmfset{arrow_len}{3mm}
\fmfset{decor_size}{2mm}
\fmfcmd{style_def majorana expr p = cdraw p; cfill (harrow (reverse p, .5)); cfill (harrow (p, .5)) enddef;
		style_def alt_majorana expr p = cdraw p; cfill (tarrow (reverse p, .55)); cfill (tarrow (p, .55)) enddef;}
\begin{alignat}{3}
	 &\Delta_{H\tH}\left(k,\omega\right)&&=\frac{-1}{\imath\omega+m_{1}+g_{1}k^{2}+\mathcal{O}\left(k^{4}\right)}&&\coloneqq
		\hspace{0.5cm}	
		\parbox{20mm}{
		\begin{fmfgraph*}(40,20)
			\fmfbottomn{v}{2}
			\fmffreeze
			\fmfshift{(0.0,0.1w)}{v1}
			\fmfshift{(0.0,0.1w)}{v2}
			\fmfv{l=$H$, l.a=90, l.d=0.02w}{v1}
			\fmfv{l=$\tH$, l.a=90, l.d=0.02w}{v2}
			\fmf{plain_arrow, label=$k,,\omega$, label.side=right}{v2,v1}
		\end{fmfgraph*}
		}\label{eq:response_propagator}\\
	&\Delta_{HH}\left(k,\omega\right)&&=\frac{D}{\omega^{2}+\left(m_{1}+g_{1}k^{2}+\mathcal{O}\left(k^{4}\right)\right)^{2}}&&\coloneqq
		\hspace{0.5cm}	
		\parbox{20mm}{
		\begin{fmfgraph*}(40,20)
			\fmfbottomn{v}{2}
			\fmffreeze
			\fmfshift{(0.0,0.1w)}{v1}
			\fmfshift{(0.0,0.1w)}{v2}
			\fmfv{l=$H$, l.a=90, l.d=0.04w}{v1}
			\fmfv{l=$H$, l.a=90, l.d=0.04w}{v2}
			\fmf{alt_majorana, label=$k,,\omega$, label.side=right}{v2,v1}
		\end{fmfgraph*}
		}\label{eq:noise_propagator}
\end{alignat}
\end{fmffile}and $\Delta_{\tilde{H}\tilde{H}}=0$. Here $\iota$ denotes the imaginary
unit. The Feynman diagram representation of both propagators is
given on the \emph{r.h.s.} of equations \prettyref{eq:response_propagator}
and \prettyref{eq:noise_propagator}. Denoting with $\hat{h}$ and
$\hat{\tilde{h}}$ the Fourier transformed fields in momentum space
alone, the propagators can also be written as a function of time and
momentum as
\begin{gather}
\Delta_{\hat{h}\hat{\tilde{h}}}\left(k,t\right)=-\theta\left(t\right)\,\exp\left(-\big(m_{1}+g_{1}k^{2}\big)\,t\right)\,,\label{eq:response_propagator_momenta_time}\\
\Delta_{\hat{h}\hat{h}}\left(k,t\right)=\frac{D}{2\left(m_{1}+g_{1}k^{2}\right)}\exp\left(-\big(m_{1}+g_{1}k^{2}\big)\,\left|t\right|\right)\,.\label{eq:noise_propagator_momenta_time}
\end{gather}
 The interaction vertices are defined by higher order functional derivatives
of $S$\begin{fmffile}{tpo_int_vertex}
\fmfset{thin}{0.75pt}
\fmfset{decor_size}{4mm}
\fmfcmd{style_def wiggly_arrow expr p = cdraw (wiggly p); shrink (0.8); cfill (arrow p); endshrink; enddef;}
\fmfcmd{style_def majorana expr p = cdraw p; cfill (harrow (reverse p, .5)); cfill(harrow (p, .5)) enddef; style_def alt_majorana expr p = cdraw p; cfill (tarrow (reverse p, .55)); cfill (tarrow (p, .55)) enddef;}
\begin{align}
\frac{1}{n!}\frac{\partial S}{\partial\tH\left(k_{1},\omega_{1}\right)\partial H\left(k_{2},\omega_{2}\right)\ldots\partial H\left(k_{n+1},\omega_{n+1}\right)}\Bigg|_{H,\tH=0} &= \left(m_{n}+g_{n}k_{1}^{2}+\mathcal{O}\left(k_{1}^{2}\right)\right)\left(2\pi\right)^{d+1}\delta\left(\sum_{i=1}^{n+1}k_{i}\right)\delta\left(\sum_{i=1}^{n+1}\omega_{i}\right) \nonumber \\
& \nonumber \\
& \coloneqq \hspace{1.4cm}\parbox{10mm}{
	\begin{fmfgraph*}(60,30)
	\fmfrightn{i}{3}
	\fmfleft{o}
	\fmftop{v}
	\fmffreeze
	\fmfshift{(-0.05w,-0.25w)}{v}
	\fmfdot{v}
	\fmflabel{$k_1,\omega_1$}{o}
	\fmflabel{$k_2,\omega_2$}{i3}
	\fmflabel{$\vdots$}{i2}
	\fmflabel{$k_{n+1},\omega_{n+1}$}{i1}
	\fmf{wiggly_arrow, tension=1.0}{v,o}
	\fmf{wiggly_arrow, tension=1.0}{i1,v}
	\fmf{wiggly_arrow, tension=1.0}{i2,v}
	\fmf{wiggly_arrow, tension=1.0}{i3,v}
	\end{fmfgraph*}
	}\label{eq:interaction_vertex}\hspace{3cm},
	\end{align} 
\end{fmffile}
\\\vspace{0.2cm}where wiggly lines represent amputated legs. The couplings $m_{n}\coloneqq l_{n}-f_{n}\sum_{\pm}w_{\pm}$
and $g_{n}\coloneqq\frac{1}{2}f_{n}\sum_{\pm}\sigma_{\pm}^{2}w_{\pm}$
are defined as in the main text by the expansion of $f$ and $l$
around $h=0$, that is $f_{n}\coloneqq\frac{1}{n!}\frac{d^{n}f}{dh^{n}}|_{h=0}$
and $l_{n}\coloneqq\frac{1}{n!}\frac{d^{n}l}{dl^{n}}|_{h=0}$.

\paragraph{Non-zero mean of the field}

In the main text, we claim $\mu$, $l_{0}$ and $f_{0}$ can in all
generality be set to zero, corresponding to $\langle h\rangle=0$.
We here justify this statement. Notice the Feynman rules above are
derived expanding the action around $h=0$. In general, non-vanishing
$\mu$, $l_{0}$ and $f_{0}$ correspond to a constant input, causing
$\langle h\rangle=\mathit{const}\neq0$. It is in general possible
to expand the action around a non-vanishing constant mean $h=\langle h\rangle$.
This gives Feynman rules formally equivalent to Eqs. \eqref{eq:response_propagator}-\eqref{eq:interaction_vertex},
with the only difference that the couplings $m_{n}$ and $g_{n}$
are defined in terms of the expansion of $f$ and $l$ around $h=\langle h\rangle$,
that is $f_{n}\coloneqq\frac{1}{n!}\frac{d^{n}f}{dh^{n}}|_{h=\langle h\rangle}$
and $l_{n}\coloneqq\frac{1}{n!}\frac{d^{n}l}{dh^{n}}|_{h=\langle h\rangle}$.
The loopwise expansion of the effective action \citep{ZinnJustin96},
for example, obeys these rules and indeed allows the study of fluctuations
of the shifted field $\delta h\coloneqq h-\langle h\rangle$. In general,
the exact value of the couplings $m_{n}$ and $g_{n}$ therefore depends
on the mean $\langle h\rangle$, which in turn depends on the input
mean $\mu$. We notice, however, that this is of no concern for a
renormalization group analysis. The latter is indeed not interested
in knowing \emph{a priori} the exact value of the couplings in the
microscopic theory. Instead, experiments are used to\emph{ a posteriori}
determine the value of the effective couplings defining the effective
theory at a certain length scale. One is rather interested in knowing
the couplings' flow for any initial condition, that is for any possible
value of the couplings in the microscopic theory. In this way one
can indeed find out which of those couplings are relevant in defining
the effective theory and need to be measured. Notice that the nature
of the flow is determined by the interaction's structure of the system,
which is the same given by Eqs. \eqref{eq:response_propagator}-\eqref{eq:interaction_vertex}
regardless of the point of expansion $h=\langle h\rangle$. For this
reason, in the main text we set in all generality $\mu$, $l_{0}$
and $f_{0}$ to zero, corresponding to $\langle h\rangle=0$. 

\paragraph{State-dependent power laws}

In a Gell-Mann Low type of flow \citep{GellMann1954}\citep[Sec. V]{Wilson75_773},
the Gaussian fixed point is reached so slowly that on many, even large
length scales the effective couplings are non-zero. Also, their slow
flow can make them appear to be approximately constant over a range
of length scales. This implies that the universality of a fixed point
is in practice never reached. In particular, the universal mean-field
scaling exponents can receive small corrections, which depend on the
value taken by the flown couplings at the scale experiments are made.
Indeed, consider an experiment made on a range of scales over which
$\bar{g}_{n}\left(\ell\right)\sim\bar{g}_{n}^{*}$ for $n=2,3$ appear
as approximately constant, because of their slow flow. Then the flow
equation \prettyref{eq:flow_nu} tells us that $g_{1}\left(\ell\right)\propto\ell^{2\epsilon}$
with $2\epsilon=\frac{3}{2}\bar{g}_{3}^{*}-\bar{g}_{2}^{*2}$. To
actually keep $g_{1}\overset{!}{=}1$ at a fixed point, the mean-field
scaling exponents therefore need to modify as $\chi\to\chi+\epsilon,\zeta\to\zeta-\epsilon,z\to z-2\epsilon$.
Notice this modification is parameterized by $\epsilon$, which depends
on the value taken by the flown couplings. This is not universal and
depends on both the scale at which experiments are made and the initial
condition for the flow, that is the value of the couplings in the
microscopic theory. It is interesting to notice how, in our model,
this initial value of the couplings depends on the state of the system,
in particular on the parameters characterizing the noisy drive coming
from other brain areas. Specifically, the variance $D$ enters directly
in the definition of the dimensionless couplings $\bar{g}_{n}\coloneqq\left(D/g_{1}\right)^{\frac{n-1}{2}}\frac{g_{n}}{g_{1}}$.
The input mean $\mu$ also determines the value of $g_{n}$ by controlling
the value of the mean $\langle h\rangle$ around which the field theory
is Taylor expanded to define the couplings (see paragraph above).

\paragraph{Balance of inhibitory and excitatory connections at criticality}

As stated in the main text, the model is at a critical point for $m_{n}\coloneqq l_{n}-f_{n}\sum_{\pm}w_{\pm}=0$.
We show that this implies a balance of inhibitory and excitatory inputs
to a neuron. We can assume $l\left(h\right)\ll w_{\pm}f\left(h\right)$,
which means that the rate at which neural activity decays is much
smaller than that at which it is updated by new inputs, whether excitatory
or inhibitory. Then, at first order, the condition for criticality
can be satisfied by the vanishing of the second term alone in the
definition of $m_{n}$, that is for $\left|w_{+}\right|=\left|w_{-}\right|$.
Recall $w_{\pm}$ are the amplitudes, of opposite sign, of the normalized
Gaussians modeling the strength of inhibitory and excitatory connections
as a function of distance. Thus their equivalence in absolute value
implies that the overall (i.e. integrated over space) strength of
inhibitory and excitatory connections is balanced.

\section{Numerical simulations}

Here we give details about the numerical simulation of \prettyref{eq:SDE_mesoscopic}.
First, we present the common ground on which we build the numerical
experiments measuring the variance and couplings' flow, as well as
the computational tasks. These then have each their own explanatory
section. Finally, in the last section we provide the values of the
parameters used to produce the figures in the main text.

\begin{figure}
\begin{centering}
\includegraphics[width=0.5\textwidth]{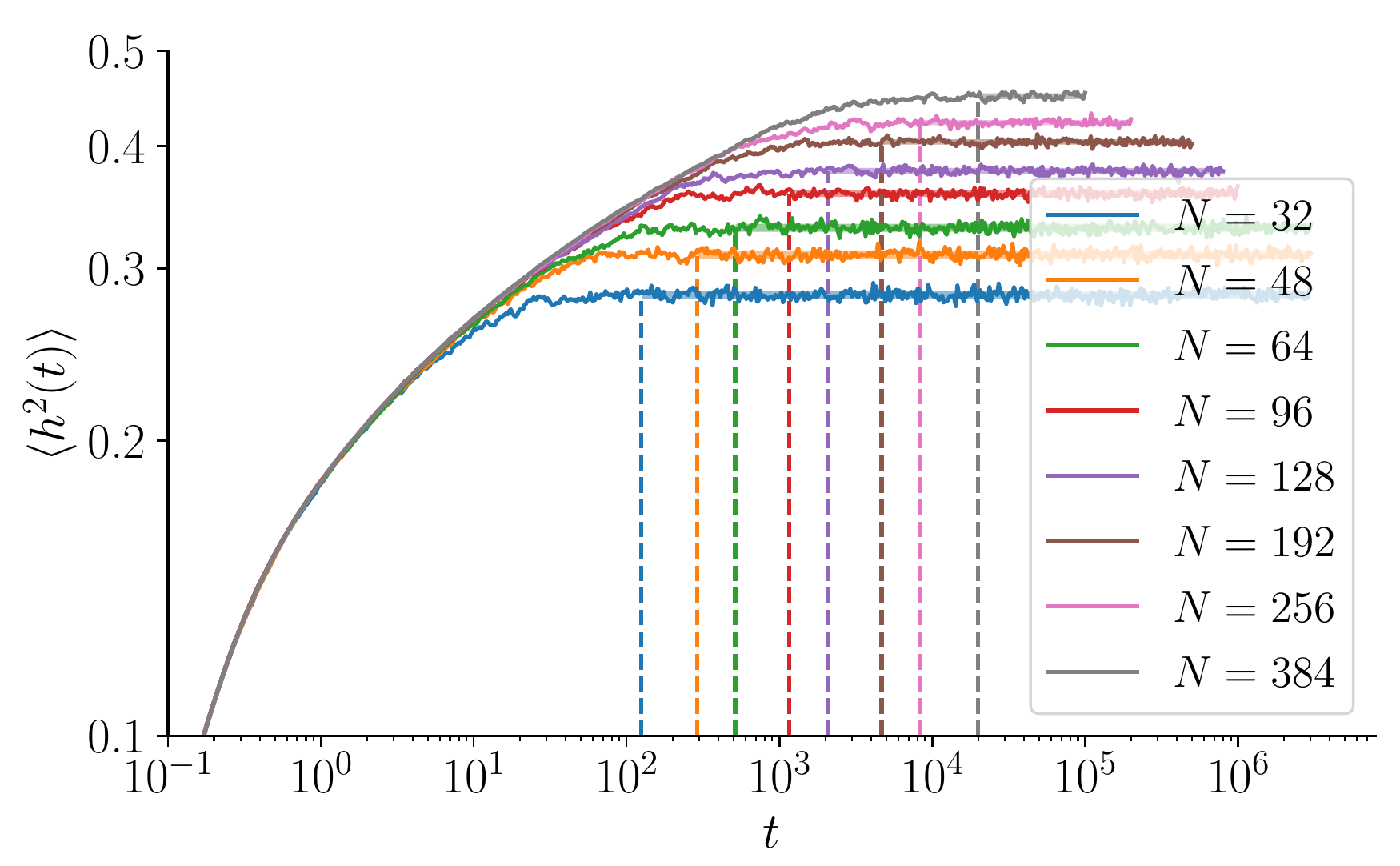}
\par\end{centering}
\caption{Measured space-averaged variance as a function of time for different
system sizes. Dashed lines denote the thermalization time $T_{\mathrm{therm}}dt$.
Each curve represents an average over $256$ trials.\label{fig:variance_therm}}
\end{figure}

Numerical simulations are performed integrating a discrete version
of \prettyref{eq:SDE_mesoscopic} with the Euler-Maruyama algorithm
\citep{Kloeden92}. The equation is simulated in its dimensionless
form, thus with $\tau=g_{1}=D=a=1$. The field variable $h\left(x,t\right)$
is defined on a square lattice of $N\times N$ sites at space points
$x=x_{1}\hat{e}_{1}+x_{2}\hat{e}_{2}$, with $x_{1},x_{2}\in\{1,\dots,N\}$
and periodic boundary conditions, and at discrete time points $t=n\,dt$,
$n\in\{0,\dots,T_{\mathrm{tot}}\}$, with $T_{\mathrm{tot}}$ the
total number of simulated time steps. The discrete equation has the
form
\begin{equation}
h\left(x,t+dt\right)-h\left(x,t\right)=-m_{1}h\left(x,t\right)\,dt+\Delta_{x,t}\left(g_{1}h+g_{2}h^{2}+g_{3}h^{3}\right)\,dt+\xi\left(x,t\right)\label{eq:discrete_SDE}
\end{equation}
where, at each time $t$ and any point $x$, the noise $\xi\left(x,t\right)$
is independently drawn from a centered Gaussian distribution with
variance $dt$. The operator $\Delta_{x,t}$ acting on a function
$f$ is defined by
\begin{equation}
\Delta_{x,t}f\coloneqq\sum_{i=1,2}\left[2f\left(x,t\right)-f\left(x+\hat{e}_{i},t\right)-f\left(x-\hat{e}_{i},t\right)\right],\label{eq:discretized_laplace}
\end{equation}
which is a common choice of discretization of the Laplace operator.
Statistical averages $\langle\quad\rangle$ are estimated by simulating
the system $R$ times with independent realizations of the noise and
averaging over these repetitions. The value of the time increment
$dt$ must be chosen small enough to guarantee numerical stability
of the integration algorithm. A value of $dt\leq0.01$ is used in
all simulations. The system is initialized at $h\left(x,t=0\right)=0$
and is thermalized for a time $T_{\mathrm{therm}}<T_{\mathrm{tot}}$,
after which measurements are made for a time $T\coloneqq T_{\mathrm{tot}}-T_{\mathrm{therm}}$.
The value of $T_{\mathrm{therm}}$ is established by measuring the
variance $\frac{1}{N^{2}}\sum_{x}\langle h^{2}\left(x,t\right)\rangle$
at all times $t$. The latter grows over time, as noise adds more
and more variability in the system, until it reaches a constant plateau
value, becoming time-independent and thus signaling thermalization,
see \prettyref{fig:variance_therm}.

\paragraph{System size and correlation length}

Though the critical point is at $m_{1}=0$, in a realistic setting
one must have a small, but finite mass $m_{1}$. This for example
prevents an unrealistic divergence of the variance of $h$ and an
infinite waiting time to reach thermalization. The mass implements
a smooth lower momentum cut-off and thus defines a finite correlation
length of $N_{\mathrm{eff}}=\frac{2\pi}{m_{1}^{2}}$ lattice sites,
which can be interpreted as the effective size of the system. Decreasing
$m_{1}$ increases the effective system size $N_{\mathrm{eff}}$.
Alternatively, one can set $m_{1}=0$ and implement a hard momentum
cut-off that only removes the $k=0$ mode from the dynamics. Since
the smallest available mode $k_{\mathrm{min}}$ then has $\left|k_{\mathrm{min}}\right|=\frac{2\pi}{N}$,
the correlation length is exactly $N_{\mathrm{eff}}=\frac{2\pi}{k_{\mathrm{min}}}=N$.
Thus the extent of correlations is increased by increasing the size
$N$ of the system itself. Analytic predictions can be made in both
cases and only differ by small minor corrections. We choose to present
here the latter approach, as it makes for more elegant analytic expressions.
The $k=0$ cut-off is implemented at each time step $t$ by subtracting
from the noise its spatial mean $\xi\left(x,t\right)\to\xi\left(x,t\right)-N^{-2}\sum_{x}\xi\left(x,t\right)$,
which in Fourier space corresponds to enforcing $\hat{\xi}\left(k=0,t\right)=0$.
The noise term is indeed the only one contributing to the $k=0$ mode
in \prettyref{eq:discrete_SDE}: the two remaining terms are indeed
one $\propto m_{1}=0$ and the other contains a Laplace operator,
which in Fourier space is proportional to $k^{2}$ for $k\to0$ and
thus gives no contribution at $k=0$.

\paragraph{Unphysical bi-stable regime}

Simulations are limited to the region $\bar{g}_{2}^{2}<3\bar{g}_{3}$,
as an unphysical stationary state emerges for $\bar{g}_{2}^{2}>3\bar{g}_{3}$.
This state has no physical interpretation, as it is just a consequence
of the polynomial approximation of $f$, ceasing to be monotonically
increasing for sufficiently large $|h|$, contrary to what is expected
from a gain function in neural systems. This alternative stationary
state has a checkerboard structure composed of two sublattices $a$
and $b$, where the neural state variable assumes the values $h_{a}$
and $h_{b}$, respectively. More precisely, a checkerboard structure
means that if $h\left(x\right)=h_{a}$ then $h\left(x\pm\hat{e}_{i}\right)=h_{b}$,
$\forall i=1,2$, and vice versa, exchanging $a\leftrightarrows b$.
The stationary state is defined in the absence of noise by the vanishing
of the r.h.s. of \prettyref{eq:discrete_SDE}. With $m_{1}=0$, this
gives the condition
\begin{equation}
g_{1}h_{a}+g_{2}h_{a}^{2}+g_{3}h_{a}^{3}=g_{1}h_{b}+g_{2}h_{b}^{2}+g_{3}h_{b}^{3}.\label{eq:spikes}
\end{equation}
The physical stationary state $h_{a}=h_{b}$ is always a solution
of \prettyref{eq:spikes}. Additional solutions emerge if the third
order polynomial in \prettyref{eq:spikes} is not monotonically increasing.
In this case $g_{1}h+g_{2}h^{2}+g_{3}h^{3}=\mathrm{const.}$ can have
three different solutions, allowing for $h_{a}\neq h_{b}$. This occurs
if
\[
\frac{g_{1}g_{3}}{g_{2}^{2}}<\frac{1}{3},
\]
which translates to $\bar{g}_{2}^{2}>3\bar{g}_{3}$ for the dimensionless
units. In simulations it turns out that the system is attracted to
this unphysical state making simulations in this parameter regime
meaningless.

\subsection{Variance}

From the simulated system, we measure the scaling of the variance
of the field $h$ with $N$. This is compared against our analytical
prediction, in which the flow of the coupling $g_{1}$ introduces
a logarithmic correction with respect to the linear case. \prettyref{subsec:Measurement-from-simulation}
provides the details on how the variance was measured from simulations.
\prettyref{subsec:Prediction-of-variance} presents the derivation
of the analytic prediction.

\subsubsection{Measurement from simulation\label{subsec:Measurement-from-simulation}}

From the simulated system, we measure the variance $\langle h^{2}\left(x,t\right)\rangle\left(N\right)$
for different lattice sizes $N$. This quantity is actually independent
of both $x$ and $t$, due to the spatio-temporal invariance of the
system. Thus, we are able to obtain a more refined measurement by
additionally averaging over space
\[
\langle h{}^{2}\left(t\right)\rangle\coloneqq\frac{1}{N^{2}}\sum_{x}\langle h^{2}\left(x,t\right)\rangle\,,
\]
and finally over time

\begin{equation}
\langle h{}^{2}\rangle\coloneqq\frac{1}{T}\sum_{t}\langle h^{2}\left(t\right)\rangle\,.\label{eq:time_average}
\end{equation}
To estimate the error $\sigma_{\langle h{}^{2}\rangle}$ in measuring
$\langle h{}^{2}\rangle$, we use the variance of \prettyref{eq:time_average}
and divide by the number $\bar{T}$ of statistically independent time
samples: 
\begin{equation}
\sigma_{\langle h{}^{2}\rangle}^{2}\coloneqq\left[\frac{1}{T}\sum_{t}\left(\langle h{}^{2}\left(t\right)\rangle-\langle h{}^{2}\rangle\right)^{2}\right]/\bar{T}\,.\label{eq:variance_of_mean}
\end{equation}
The number $\bar{T}$ is smaller than the total number $T$ of time
samples because two time samples can be considered statistically independent
only when they are at a distance larger than the temporal correlation
length. An estimate of this length is naturally given by the thermalization
time $T_{\mathrm{therm}}dt$, as this is the time after which the
system completely forgets any initial condition. Using a conservative
approach, we consider two time points to be independent when they
are at a distance of $3T_{\mathrm{therm}}dt$. Thus, the number of
independent time samples we use is $\bar{T}\coloneqq T/\left(3T_{\mathrm{therm}}\right)$.
\prettyref{fig:variance_therm} illustrates the measured space-averaged
variance $\langle h^{2}\left(t\right)\rangle$ and its thermalization
over time $t$.

\subsubsection{Prediction of variance by RG\label{subsec:Prediction-of-variance}}

The measured scaling of the variance with the system size $N$ is
compared against analytical predictions. The variance can be written
as
\begin{equation}
\langle h{}^{2}\rangle=\langle h{}^{2}\left(x=0,t\right)\rangle=\int\frac{d^{2}k}{\left(2\pi\right)^{2}}\int\frac{d^{2}k^{\prime}}{\left(2\pi\right)^{2}}\langle\hat{h}\left(k,t\right)\hat{h}\left(k^{\prime},t\right)\rangle\,.\label{eq:variance_momenta}
\end{equation}
We then use the scaling relation
\begin{equation}
\langle\hat{h}\left(k,t\right)\hat{h}\left(k^{\prime},t\right)\rangle=\ell^{2\zeta}\langle\hat{h}_{\ell}\left(\ell k,\ell^{-2}t\right)\hat{h}_{\ell}\left(\ell k^{\prime},\ell^{-2}t\right)\rangle_{\ell}\,,\label{eq:scaling_relation}
\end{equation}
where $\zeta=\left(d+2\right)/2\overset{d=2}{=}2$ is the scaling
exponent of the field $\hat{h}$, as defined in the main text. The
subscript $\ell$ in $\langle\rangle_{\ell}$ indicates that averaging
must be performed according to the statistics defined by the effective
theory at scale $\ell$. We choose the highest possible value of the
flow parameter $\ell=N/2$, so that all degrees of freedom have been
marginalized in the effective theory, that is those with momenta $\left|k\right|\geqslant\frac{\Lambda}{\ell}=\frac{2\pi}{N}=\left|k_{\mathrm{min}}\right|$
(recall $k=0$ is removed from the dynamics). With this choice, no
loop diagrams, which account for statistical fluctuations, contribute
to the \emph{r.h.s. }of \prettyref{eq:scaling_relation}. This means
only the tree level diagram of the form found in \prettyref{eq:noise_propagator}
remains and thus
\[
\ell^{2\zeta}\langle\hat{h}_{\ell}\left(\ell k,\ell t\right)\hat{h}_{\ell}\left(\ell k^{\prime},\ell t\right)\rangle_{\ell}=\ell^{2}\Delta_{\hat{h}_{\ell}\hat{h}_{\ell}}\left(\ell k,0\right)\delta\left(k+k^{\prime}\right)=\frac{1}{2g_{1}\left(\ell\right)k^{2}}\delta\left(k+k^{\prime}\right)\,.
\]
The effect of fluctuations is hidden in the flown coupling $g_{1}\left(\ell=N/2\right)$.
With this information, \prettyref{eq:variance_momenta} becomes
\begin{equation}
\langle h{}^{2}\rangle=\int\frac{d^{2}k}{\left(2\pi\right)^{2}}\ell^{2}\Delta_{\hat{h}_{\ell}\hat{h}_{\ell}}\left(\ell k,0\right)\bigg|_{\ell=\frac{N}{2}}=\frac{1}{4\pi}\int_{\frac{2\pi}{N}}^{\,\pi}\frac{kdk}{g_{1}\left(N/2\right)k^{2}}=\frac{1}{4\pi}\frac{\ln\left(N/2\right)}{g_{1}\left(N/2\right)}\,.\label{eq:variance_scaling_continuous}
\end{equation}
In the second step, the angular part of the momentum integral is performed.
Notice that the remaining radial integral starts from the lowest available
mode $\left|k_{\mathit{min}}\right|=\frac{2\pi}{N}$, thus giving
the dependence of the variance on the correlation length (i.e. the
system size $N$).

\prettyref{eq:variance_scaling_continuous} is strictly valid only
in the limit of an infinite-size system, $N\to\infty$. Simulations
are, however, restricted to finite values of $N$. To make a prediction
that compares accurately with the simulated finite-sized systems,
we need to apply some corrections to \prettyref{eq:variance_scaling_continuous}.
First, for finite $N,$ the Fourier transformed fields $\hat{h}$
are defined only on a discrete, rather than continuous, set of momenta.
These are $k=k_{1}\hat{e}_{1}+k_{2}\hat{e}_{2}$, with $k_{i}=\frac{2\pi}{N}n_{i}$
and $n_{i}\in\left(\frac{N}{2}-1,\frac{N}{2}\right)\cap\mathbb{Z}$.
Thus the momentum integral in \prettyref{eq:variance_scaling_continuous}
must be substituted with the sum 
\begin{equation}
\frac{1}{N^{2}}\sum_{k}\overset{N\to\infty}{\to}\int\frac{d^{2}k}{\left(2\pi\right)^{2}}.\label{eq:sum_integral}
\end{equation}
Notice that the sum reduces to the integral when $N\to\infty$. Secondly,
the propagator in \prettyref{eq:variance_scaling_continuous} must
be substituted with its analogue for finite-size systems. This can
be derived from the expression for the infinite-size system's propagator
\prettyref{eq:noise_propagator}. Notice that the $k^{2}$ appearing
in the expression arises from the Fourier representation of the Laplace
operator. Thus, one simply needs to substitute $k^{2}$ with the Fourier
representation of the discrete Laplace operator \prettyref{eq:discretized_laplace},
which is $\sum_{i}\left(2\cos\left(k_{i}\right)-2\right)$. The finite-size
propagator thus has the form
\begin{equation}
\Delta_{\hat{h}\hat{h}}\left(k,0\right)^{-1}=2g_{1}\sum_{i}\left(2\cos\left(k_{i}\right)-2\right)\overset{k\to0}{\to}2g_{1}k^{2}\,.\label{eq:discrete_inverse_propagator}
\end{equation}
Notice that the finite-size propagator reduces to the infinite-size
one in the limit of large wave-lengths $k\to0$. In \prettyref{eq:variance_scaling_continuous},
the propagator is that of the effective theory at scale $\ell=N/2$.
This means the couplings appearing in the propagator must be evaluated
at the point $\ell=N/2$ of their flow. In the infinite-size propagator,
the only coupling appearing is $g_{1}$; but the finite-size propagator
contains additional correction terms, each with its own coupling.
We make these explicit by Taylor expanding the difference between
the finite and infinite-size propagator
\[
\Delta_{\hat{h}\hat{h}}\left(k,0\right)^{-1}-2g_{1}k^{2}\coloneqq\sum_{n=4}^{\infty}c_{n}\sum_{i}k_{i}^{n}\,,
\]
with the expansion coefficients $c_{n}$ being the additional couplings.
These are all of order $\mathcal{O}\left(k^{4}\right)$ and thus irrelevant:
mean-field (i.e. dimensional) analysis predicts their flow to be $c_{n}\left(\ell\right)=\ell^{-\left(n-2\right)}c_{n}\left(0\right)$.
We can assume this fast power-law decay to dominate their flow, and
thus neglect higher order loop corrections. With this approximation,
the flown propagator is given by
\begin{align}
\ell^{-2}\Delta_{\hat{h}_{\ell}\hat{h}_{\ell}}\left(\ell k,0\right)^{-1} & =2g_{1}\left(\ell\right)k^{2}+\sum_{n=4}^{\infty}c_{n}\left(\ell\right)\sum_{i}\ell^{n-2}k_{i}^{n}\nonumber \\
 & =2g_{1}\left(\ell\right)k^{2}+\sum_{n=4}^{\infty}c_{n}\left(0\right)\sum_{i}k_{i}^{n}\nonumber \\
 & =2g_{1}\left(\ell\right)k^{2}+2g_{1}\left(0\right)\left[\sum_{i}\left(2\cos\left(k_{i}\right)-2\right)-k^{2}\right].\label{eq:discrete_flown_propagator}
\end{align}
We can finally plug the modifications Eqs. \eqref{eq:sum_integral}
and \eqref{eq:discrete_flown_propagator} into \prettyref{eq:variance_scaling_continuous}
to obtain a finite-size system's prediction for the variance 
\begin{equation}
\langle h{}^{2}\rangle=\frac{1}{2N^{2}}\sum_{k\neq0}\frac{1}{g_{1}\left(N/2\right)k^{2}+g_{1}\left(0\right)\left[\sum_{i}\left(2\cos\left(k_{i}\right)-2\right)-k^{2}\right]}\,.\label{eq:discrete_variance}
\end{equation}
The sum over discrete momenta appearing in \prettyref{eq:discrete_variance}
is carried out numerically.

\subsection{Flow of the couplings}

From the simulated system we also measure $n$-point correlations
in momentum space up to $n=4$, as a function of the correlation length
(i.e. system size) $N$. These can be expressed in terms of the couplings
$g_{n}\left(\ell\right)$ at $\ell=N/2$. This allows us, by varying
the simulated system's size, to obtain a measured value of the couplings
at different points of their flow. In \prettyref{subsec:Relationship-between-couplings}
we derive the analytic expression relating correlations to the flown
couplings. In \prettyref{subsec:Measurement-of-flown} we detail how
correlations are measured from simulations and the analytic expressions
inverted to extract a measured value of the flown couplings.

\subsubsection{Relationship between couplings and correlations\label{subsec:Relationship-between-couplings}}

With the same reasoning as for the derivation of the variance \prettyref{subsec:Prediction-of-variance},
we make use of the scaling law 
\begin{equation}
\langle\langle\hat{h}\left(k_{1},t\right)\ldots\hat{h}\left(k_{n},t\right)\rangle\rangle=\ell^{n\zeta}\langle\langle\hat{h}_{\ell}\left(\ell k_{1},\ell^{-2}t\right)\ldots\hat{h}_{\ell}\left(\ell k_{n},\ell^{-2}t\right)\rangle\rangle_{\ell}\,,\label{eq:scaling_cumulants}
\end{equation}
where $\zeta=\left(d+2\right)/2\overset{d=2}{=}2$ is the scaling
exponent of the field $\hat{h}$, as defined in the main text, and
$\langle\langle\,\rangle\rangle$ denotes cumulants, rather than moments.
Evaluating at $\ell=N/2$, we only have to compute the r.h.s. of \prettyref{eq:scaling_cumulants}
for a theory where all fluctuations have been integrated out. This
means we get contributions from tree-level diagrams alone, and no
loop diagrams, which would correspond to fluctuations. The diagrams
contributing to the $n=2,3,4$-point correlation functions are respectively
of the form\begin{fmffile}{tpo_app_moments}
\fmfset{thin}{0.75pt}
\fmfset{arrow_len}{3mm}
\fmfcmd{style_def majorana expr p = cdraw p; cfill (harrow (reverse p, .5)); cfill (harrow (p, .5)) enddef;
		style_def alt_majorana expr p = cdraw p; cfill (tarrow (reverse p, .55)); cfill (tarrow (p, .55)) enddef;}
\begin{equation}
\begin{aligned}
	 & \hspace{0.5cm}	
		\parbox{30mm}{
		\begin{fmfgraph*}(70,50)
			\fmfbottomn{v}{2}
			\fmffreeze
			\fmfshift{(0.0,0.40h)}{v1}
			\fmfshift{(0.0,0.40h)}{v2}
			\fmf{alt_majorana}{v2,v1}
		\end{fmfgraph*}
		}; & \hspace{0.5cm}	
		\parbox{30mm}{
		\begin{fmfgraph*}(60,50)
			\fmfstraight
			\fmfright{v2,v3}
			\fmfbottom{v1,c}
			\fmffreeze
			\fmfshift{(0.0w,0.5h)}{v1}
			\fmfshift{(-0.5w,0.5h)}{c}
			\fmfdot{c}
			\fmf{plain_arrow}{c,v1}
			\fmf{alt_majorana}{c,v2}
			\fmf{alt_majorana}{c,v3}
		\end{fmfgraph*}
		}; & \\
	& \hspace{0.5cm}	
		\parbox{30mm}{
		\begin{fmfgraph*}(70,50)
			\fmfstraight
			\fmfright{v2,v3,v4}
			\fmfbottom{v1,c}
			\fmffreeze
			\fmfshift{(0.0w,0.5h)}{v1}
			\fmfshift{(-0.6w,0.5h)}{c}
			\fmfshift{(-0.1w,0.0h)}{v2}
			\fmfshift{(-0.1w,0.0h)}{v4}
			\fmfdot{c}
			\fmf{plain_arrow}{c,v1}
			\fmf{alt_majorana}{c,v2}
			\fmf{alt_majorana}{c,v3}
			\fmf{alt_majorana}{c,v4}
		\end{fmfgraph*}
		} & \hspace{0.25cm}	
		\parbox{30mm}{
		\begin{fmfgraph*}(80,50)
			\fmfstraight
			\fmfleft{v1,v2,v3}
			\fmfright{v4,v5,v6}
			\fmffreeze
			\fmfshift{(0.25w,0.0h)}{v2}
			\fmfshift{(-0.25w,0.0h)}{v5}
			\fmfdot{v2,v5}
			\fmf{plain_arrow}{v2,v1}
			\fmf{alt_majorana}{v2,v3}
			\fmf{alt_majorana}{v5,v4}
			\fmf{alt_majorana}{v5,v6}
			\fmf{plain_arrow}{v5,v2}
		\end{fmfgraph*}
		} &
		\hspace{1.0cm}	
		\parbox{30mm}{
		\begin{fmfgraph*}(80,50)
			\fmfstraight
			\fmfleft{v1,v2,v3}
			\fmfright{v4,v5,v6}
			\fmffreeze
			\fmfshift{(0.25w,0.0h)}{v2}
			\fmfshift{(-0.25w,0.0h)}{v5}
			\fmfdot{v2,v5}
			\fmf{plain_arrow}{v2,v1}
			\fmf{alt_majorana}{v2,v3}
			\fmf{plain_arrow}{v5,v4}
			\fmf{alt_majorana}{v5,v6}
			\fmf{alt_majorana}{v5,v2}
		\end{fmfgraph*}
		}\label{eq:cumulant_234_diag}.
\end{aligned}
\end{equation}
\end{fmffile}Computed explicitly, these diagrams give the results
\begin{equation}
\langle\langle\hat{h}\left(k_{1},t\right)\hat{h}\left(k_{2},t\right)\rangle\rangle=\frac{D}{2g_{1}\left(N/2\right)k_{1}^{2}}\delta\left(k_{1}+k_{2}\right),\label{eq:2_cumulant}
\end{equation}
\begin{equation}
\langle\langle\hat{h}\left(k_{1},t\right)\hat{h}\left(k_{2},t\right)\hat{h}\left(k_{3},t\right)\rangle\rangle=-\frac{1}{2}\left(\frac{D}{g_{1}\left(N/2\right)}\right)^{2}\frac{g_{2}\left(N/2\right)}{g_{1}\left(N/2\right)}\frac{\delta\left(k_{1}+k_{2}+k_{3}\right)}{k_{1}^{2}+k_{2}^{2}+k_{3}^{2}}\frac{k_{1}^{2}}{k_{2}^{2}k_{3}^{2}}+\mathit{perms},\label{eq:3_cumulant}
\end{equation}
\begin{align}
\langle\langle\hat{h}\left(k_{1},t\right)\ldots\hat{h}\left(k_{4},t\right)\rangle\rangle=\left(\frac{D}{g_{1}\left(N/2\right)}\right)^{3} & \frac{\delta\left(k_{1}+k_{2}+k_{3}+k_{4}\right)}{k_{1}^{2}+k_{2}^{2}+k_{3}^{2}+k_{4}^{2}}\Bigg(-\frac{1}{8}\frac{g_{3}\left(N/2\right)}{g_{1}\left(N/2\right)}\frac{k_{1}^{2}}{k_{2}^{2}k_{3}^{2}k_{4}^{2}}\nonumber \\
 & +\frac{1}{4}\left(\frac{g_{3}\left(N/2\right)}{g_{1}\left(N/2\right)}\right)^{2}\frac{\left(k_{1}+k_{2}\right)^{2}k_{4}^{2}}{k_{1}^{2}k_{2}^{2}k_{3}^{2}\left(\left(k_{1}+k_{2}\right)^{2}+k_{4}^{2}+k_{2}^{2}\right)}\nonumber \\
 & +\frac{1}{4}\left(\frac{g_{3}\left(N/2\right)}{g_{1}\left(N/2\right)}\right)^{2}\frac{k_{2}^{2}k_{4}^{2}}{k_{1}^{2}k_{3}^{2}\left(k_{1}+k_{2}\right)^{2}\left(\left(k_{1}+k_{2}\right)^{2}+k_{1}^{2}+k_{2}^{2}\right)}\Bigg)+\mathit{perms}.\label{eq:4_cumulant}
\end{align}
The notation $+\mathit{perms}$ means that the result is actually
given by the sum of all permutations of momenta $k_{i}$ in the expressions
on the r.h.s of Eqs. \eqref{eq:3_cumulant} and \eqref{eq:4_cumulant}.
This corresponds to all possible ways of assigning momenta to the
external legs of the diagrams in \prettyref{eq:cumulant_234_diag}.

\subsubsection{Measurement of flown couplings from simulation\label{subsec:Measurement-of-flown}}

The measurement of the couplings $g_{n}\left(N/2\right)$ from the
simulation follows the procedure described below. For each simulated
time step $t$, the discrete Fourier transform $\hat{h}\left(t\right)$
of the measured state $h\left(t\right)$ is computed. Then, trial
averaging is performed to estimate the cumulants on the l.h.s of Eqs.
\eqref{eq:2_cumulant} to \eqref{eq:4_cumulant}. These are only measured
for momenta combinations satisfying the momentum conservation law
$\sum_{n}k_{n}=0$. For these combinations, the Dirac delta appearing
on the r.h.s. of Eqs. \eqref{eq:2_cumulant} to \eqref{eq:4_cumulant}
takes the value $\delta\left(0\right)=N^{2}$ (recall we consider
a discrete system, so $\delta\left(0\right)\to\infty$ only for $N\to\infty$).
Then, the values of $g_{n}\left(N/2\right)$ for $n=1,2,3$ are extracted
one after the other by inverting Eqs. \eqref{eq:2_cumulant}, \eqref{eq:3_cumulant}
and \eqref{eq:4_cumulant}. This gives a measured value of $g_{n}\left(N/2\right)$
for each momenta combination and time step. The flown couplings are,
however, independent of these parameters so an additional averaging
over momenta and time is performed. At the stage of time averaging,
which is performed last, the statistical error is estimated, in the
same way as is done for the variance measurement \prettyref{subsec:Measurement-from-simulation}.
Finally, we note that cumulants are only measured for small momenta
$\left|k\right|<k_{\mathit{max}}=0.44$. This is done because Eqs.
\eqref{eq:2_cumulant} to \eqref{eq:4_cumulant} are only valid in
the continuum limit $k\to0$. In particular, the chosen value of $k_{\mathit{max}}$
guarantees that the relative error between the discrete and continuous
theory propagators, Eqs. \eqref{eq:discrete_inverse_propagator} and
\eqref{eq:noise_propagator_momenta_time}, is at most $1.7\%$ for
the largest momenta.

\subsection{Memory task\label{subsec:Memory-task}}

The simulated model is used as a reservoir computing system to perform
a mnemonic task. The task is designed to measure the decay in the
memory of a stimulus that is presented to the system.

The basic protocol comprises the following steps. First, the system
evolves until time $t_{\mathrm{in}}=T_{\mathrm{therm}}$, reaching
a thermalized state $h_{0}\left(x\right)\coloneqq h\left(x,t_{\mathrm{in}}\right)$.
Second, at time $t_{\mathrm{in}}$ a stimulus $s\left(x\right)$ consisting
of a two-dimensional Gaussian shape is added on top of the thermalized
state, $h\left(x,t_{\mathrm{in}}\right)\to h\left(x,t_{\mathrm{in}}\right)+s\left(x\right)$.
The perturbed state $h\left(x,t\right)$ evolves further in time,
still following \prettyref{eq:discrete_SDE}. Finally, at readout
times $\left\{ t_{i}|i=1,\ldots,n\right\} $ we take snapshots $h\left(x,t_{i}\right)$
and measure the accuracy $F\left(t_{i}\right)$ with which a trained
linear readout can reconstruct the stimulus from $h\left(x,t_{i}\right)$.
The reconstructed stimulus at time $t_{i}$ is defined as
\[
z\left(x,t_{i}\right)\coloneqq\sum_{x'}W^{\left(i\right)}\left(x,x'\right)\,h\left(x',t_{i}\right),
\]
where $W^{\left(i\right)}$ is an $N^{2}\times N^{2}$ readout weight
matrix, whose entries $W^{\left(i\right)}\left(x,x'\right)$ map a
lattice point $x'$ of the simulated system to a lattice point $x$
of the reconstructed stimulus, cf. \prettyref{fig:parity_labels}(a).
We define the reconstruction error $E$, called \textit{loss}, as
the spatial mean over the squared difference between the stimulus
and its reconstruction
\[
E\left(t_{i}\right)\coloneqq\frac{1}{N^{2}}\sum_{x}\left[z\left(x,t_{i}\right)-s\left(x\right)\right]^{2},
\]
which is equal to zero for perfect retrieval of the stimulus. Hence,
$F\left(t_{i}\right)=1-E\left(t_{i}\right)/\mathcal{N}_{i}$, with
$\mathcal{N}_{i}$ some normalization factor, is a convenient measure
for the reconstruction accuracy, as it has a maximum of $1$ when
the stimulus is perfectly reconstructed. The normalization factor
is chosen such that the reconstruction accuracy is on average $0$
when the stimulus has been completely forgotten by the system. This
situation corresponds to the field $h\left(x,t_{i}\right)$ obeying
the statistics of the thermalized, yet unperturbed state $h_{0}\left(x\right)$.
In this case we have 
\begin{gather}
\langle E\left(t_{i}\right)\rangle=\frac{1}{N^{2}}\sum_{x}\left[\langle z_{0}^{2}\left(x,t_{i}\right)\rangle+s^{2}\left(x\right)\right]\eqqcolon\mathcal{N}_{i}\label{eq:fidelity_normalization}\\
\mathrm{with}\quad z_{0}\left(x,t_{i}\right)\coloneqq\sum_{x'}W^{\left(i\right)}\left(x,x'\right)\,h_{0}\left(x',t_{i}\right),\nonumber 
\end{gather}
where we used that $\langle z_{0}\rangle=0$. The statistical average
appearing in \prettyref{eq:fidelity_normalization} is estimated within
the training procedure outlined below. This indeed produces independent
copies of the thermalized state $h_{0}$ which are used as trials
to estimate $\langle z_{0}^{2}\rangle$.

\paragraph{Training the readout}

We use gradient descent \citep{Kiefer52_462} to train the readout
weight matrices $W_{i}$. To this end, we repeat the above protocol
several times with different realizations of the noise $I$ and in
each iteration the weights are adjusted to minimize the loss. The
gradient of the loss function with respect to the weights reads
\[
\frac{\partial E\left(t_{i}\right)}{\partial W^{\left(i\right)}\left(x,x'\right)}=\frac{2}{N^{2}}\left[z\left(x,t_{i}\right)-s\left(x\right)\right]\,h\left(x',t_{i}\right).
\]
Therefore, the weight increment at each training step can be expressed
as $\Delta W^{\left(i\right)}\left(x,x'\right)=-\eta\left[z\left(x,t_{i}\right)-s\left(x\right)\right]\,h\left(x',t_{i}\right)$.
All prefactors are absorbed into the adjustable learning rate $\eta$
which has to be tuned such that the weights converge during the training.
To achieve a faster and more robust convergence, we employ batch learning.
One batch consists of multiple copies of the system and the readout
weights. The copies are simulated in parallel with different noise
realizations. After each training iteration, the gradient is computed
for each copy and averaged over the batch. The averaged gradient is
then used to update the weights. At the end of the training, the reconstuction
accuracy of the last iteration, averaged over the batch, is stored.
Batch averaging is also used to estimate $\langle z_{0}{}^{2}\rangle$,
which appears in the definition of the accuracy's normalization factor
\prettyref{eq:fidelity_normalization}. The whole training process
is repeated several times with different noise realizations. The achieved
accuracy's mean and standard deviation over these trials is computed.

\subsection{Classification task}

The simulated model is used as a reservoir computing system to perform
a nonlinear classification task.

The task exploits the system's ability to perform nonlinear transformations
to implement the parity function $p:\left\{ -1,1\right\} ^{\alpha}\rightarrow\left\{ 0,1\right\} $.
It maps the binary representation $v\in\left\{ -1,1\right\} ^{\alpha}$
of a given number to $p\left(\nu\right)=1$ if and only if the number
of ones in the input $v$ is odd, and to $p\left(v\right)=0$ otherwise.
For the task presented in \prettyref{fig:Examples-of-computation.}(b),
we use $\alpha=3$. The labeling according to the parity function
is linearly non-separable, see main text \prettyref{fig:Examples-of-computation.}(c).

\begin{figure}
\begin{centering}
\includegraphics[width=0.4\textwidth]{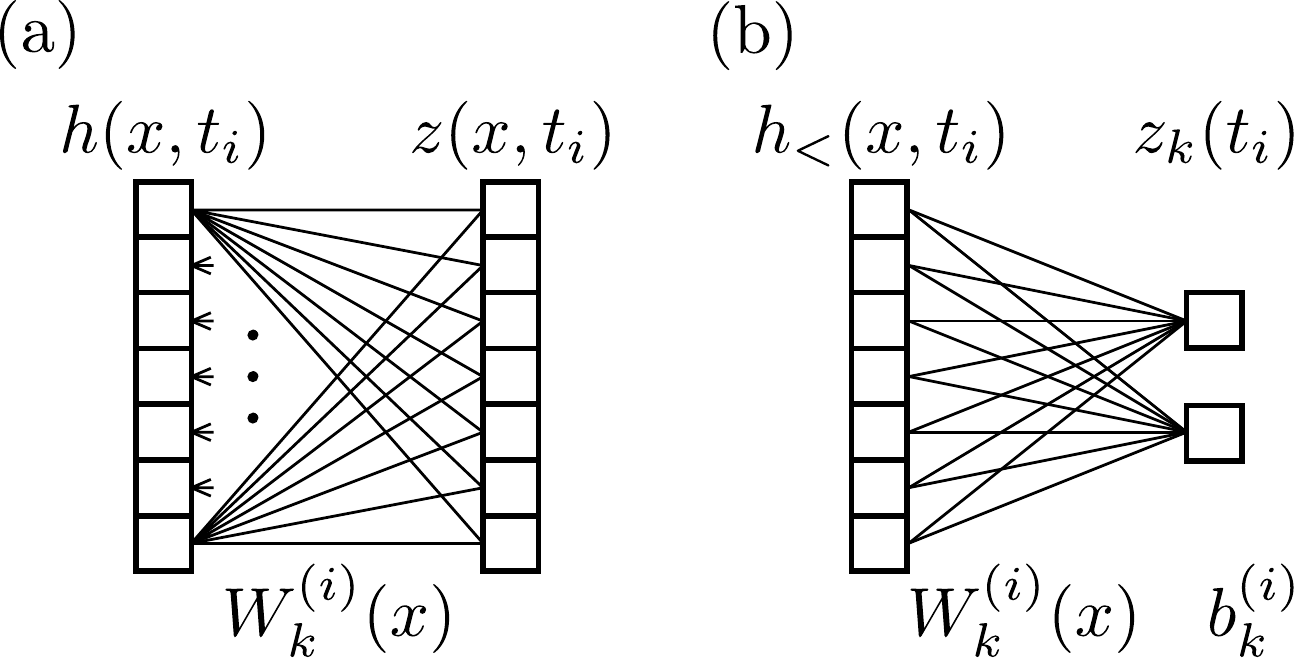}
\par\end{centering}
\caption{(a)\textbf{ }Sketch of the readout at time $t_{i}$ for the memory
task. (b) Sketch of the readout at time $t_{i}$ for the classification
task.\label{fig:parity_labels}}
\end{figure}

The protocol of this task is similar to that of the memory task. The
system evolves until a stationary state is reached, then a stimulus
is added which encodes a binary representation $v$ in a pattern consisting
only of frequencies greater than a cutoff frequency $k_{\mathrm{cut}}$.
In our case we use $k_{\mathrm{cut}}=\pi/2$. Details of the encoding
are discussed below. Finally we use trained linear readouts at different
times $t_{i}$ to predict the parity of the stimulus. The readouts
act on low-pass filtered snapshots $h_{<}\left(x,t_{i}\right)$. The
snapshots $h_{<}\left(x,t_{i}\right)$ contain only frequencies smaller
than $k_{\mathrm{cut}}$ so that the original stimulus is excluded
and only the emerging lower modes are kept. This is done to further
emphasize the role of nonlinearities in solving the task: not only
are they needed to realize the mapping $p$, but also to transfer
information from the higher modes where the input is fed to the lower
modes from which the output is read out. The latter task is unsolvable
for a linear system as modes are independent, which means that they
cannot interact to transfer the information. For each $t_{i}$ there
are two readout units $z_{k}\left(t_{i}\right)$, $k=1,2$, defined
by
\[
z_{k}\left(t_{i}\right)=b_{k}^{\left(i\right)}+\sum_{x}W_{k}^{\left(i\right)}\left(x\right)\,h_{<}\left(x,t_{i}\right),
\]
where the linear readout is performed by the weights $W_{k}^{\left(i\right)}\left(x\right)$,
mapping a lattice point $x$ to the readout unit $k$, and the biases
$b_{k}^{\left(i\right)}$, causing an overall shift in the readout
unit $k$, cf. \prettyref{fig:parity_labels}(b). The two units represent
the two classes of even ($k=1$) and odd ($k=2$) parity. Correspondingly,
if $z_{1}\left(t_{i}\right)>z_{2}\left(t_{i}\right)$, the system's
prediction is assumed to be ``even'', while it is ``odd'' in the
opposite case. By repeating this procedure for all possible inputs
several times with different noise realizations we compute the share
of correctly classified inputs. As in the other task, we also perform
the training several times and average over the results to get the
mean and variance of the classification performance.
\begin{figure}
\begin{centering}
\includegraphics[width=0.8\textwidth]{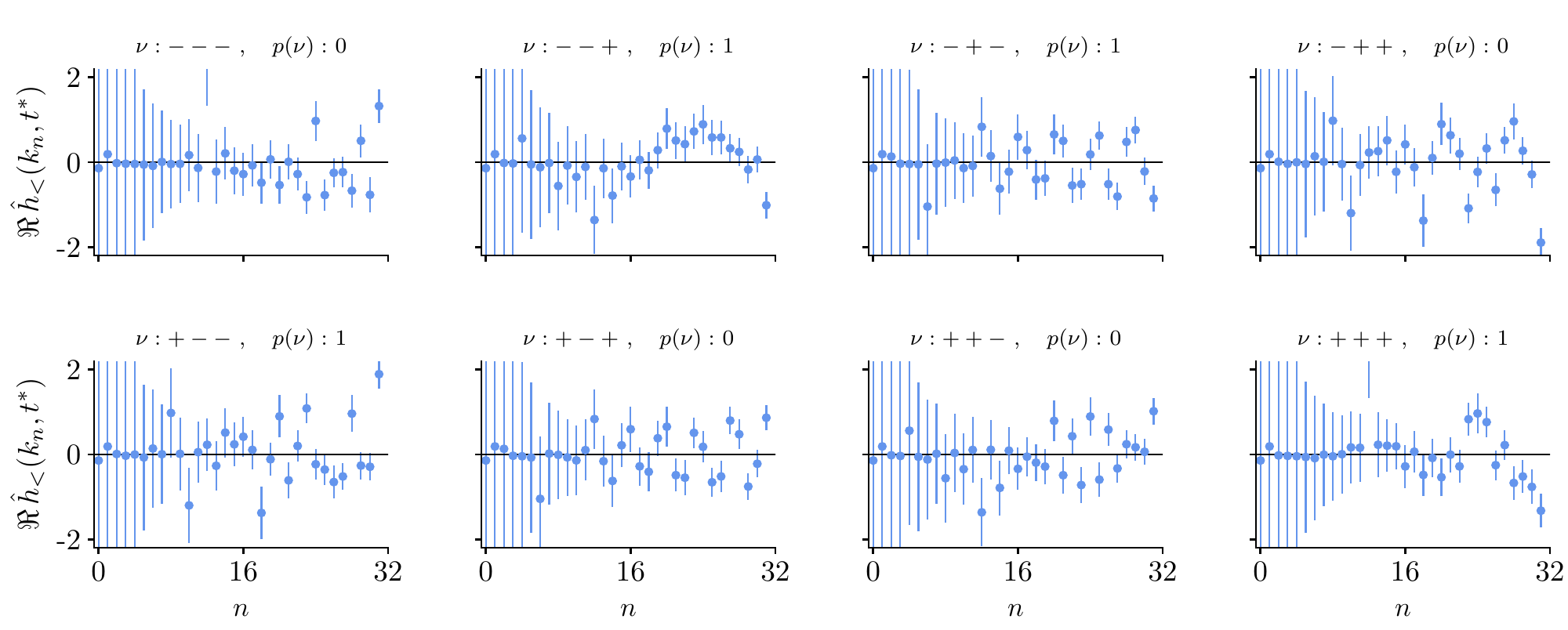}
\par\end{centering}
\caption{Diversity of patterns in the system's modes response to input stimuli.
The Fourier transform $\hat{h}_{<}\left(k,t^{*}\right)$ of the low-pass
filtered snapshots $h_{<}\left(x,t^{*}\right)$ used to train the
readout in the classification task \prettyref{fig:Examples-of-computation.}(b)
is represented. The system size is $N=128$ and $k_{\mathrm{cut}}=\pi/2$;
so only the modes $k_{n}=2\pi n/128$, with $n=0,\ldots,31$, pass
the filter. Dots represent the real part of the Fourier transform
along the $x_{1}$ direction averaged over 4096 trials (the stimulus
is constant in $x_{2}$ direction so that this component does not
carry any information). The vertical bars represent the standard deviation.
The different panels show the emerging distribution of momentum modes
in response to the eight different $3$-bit strings. The readout time
after input injection is $\Delta t=0.5$.\label{fig:k_modes}}
\end{figure}

\paragraph{Representation of the input}

Each input $\nu$ is represented by a different stimulus $s_{\nu}\left(x\right)$.
Each of its bit components $\nu_{n}\in\left\{ -1,1\right\} $, $n=1,\ldots,\alpha$,
is represented by a spatial oscillation in the $x_{1}$ direction
of frequency $\pi-n\,\delta k$ and amplitude $\nu_{n}$, which are
superimposed to give

\[
s_{\nu}\left(x\right)=A\sum_{n=1}^{\alpha}\nu_{n}\cos\left(\left(\pi-n\,\delta k\right)x_{1}\right),
\]
with $A$ some overall signal amplitude. The distance between the
excited modes is given by $\delta k=2\frac{2\pi}{N}$. Notice that,
with $N=128$ and $\alpha=3$ as in \prettyref{fig:Examples-of-computation.}(b),
all stimuli contain only modes higher than the cutoff frequency $k_{\mathrm{cut}}=\pi/2$.

\paragraph{Training the readout}

Again, we use gradient descent \citep{Kiefer52_462} to train the
weights and biases. The loss function is the cross entropy error \citep{Hinton89_185}
defined by
\[
E\left(t_{i}\right)=-\sum_{k}y_{k}^{\ast}\ln y_{k}\left(t_{i}\right),
\]
where $y_{k}\left(t_{i}\right)$, $k=1,2$, is computed as the softmax
of the bare output
\[
y_{k}\left(t_{i}\right)=\frac{e^{z_{k}\left(t_{i}\right)}}{\sum_{k}e^{z_{k}\left(t_{i}\right)}}.
\]
The correct label is one-hot encoded in $y^{\ast}$, which means $y^{\ast}=\left(1,0\right)^{\mathrm{T}}$
if the input's parity is even and $y^{\ast}=\left(0,1\right)^{\mathrm{T}}$
otherwise. The gradients with respect to the biases and weights are
given by
\begin{align*}
\frac{\partial E}{\partial W_{kl}^{\left(i\right)}}\left(t_{i}\right) & =\left(y_{k}\left(t_{i}\right)-y_{k}^{\ast}\right)\,h_{<}\left(t_{i},x_{l}\right),\\
\frac{\partial E}{\partial b_{k}^{\left(i\right)}}\left(t_{i}\right) & =\left(y_{k}\left(t_{i}\right)-y_{k}^{\ast}\right).
\end{align*}
Also in this task we employ batch learning to make the training more
robust.

\paragraph{Response of readout modes to stimuli}

In \prettyref{fig:k_modes} we show the Fourier modes' response to
the $8$ different $3$-bit strings. This shows the system's modes
are interacting, with a degree of nonlinearity sufficient to perform
the task. First, notice that all the lower modes that are used at
readout respond to the input, which is originally fed to higher modes
only. This requires interaction between modes to transfer information
of the input, which is not possible in a linear system. Secondly,
notice the diversity of patterns produced in response to the different
$3$-bit strings. This realizes the dimensionality expansion necessary
to perform the task: from initially three dimensions, nonlinear interactions
spread the inputs in the higher dimensional space of readout modes,
such that a hyperplane in that space can be found that separates them
correctly according to their parity.

\subsection{Simulation parameters}

\paragraph*{Figure 1}

The parameters of the simulations are the same for \prettyref{fig:Phase-diagram}(b),
(c), and (d) and depend only on the system size $N$. They are summarized
in \prettyref{tab:fig1}. The last row contains the order of magnitude
of statistically independent measurements $R\cdot\bar{T}$ (recall
$\bar{T}$ is the number of statistically independent time steps;
$R$ is the number of repeated simulations of the system, with an
independent realization of the noise). In all simulations $dt=0.01$.
The initial values in panel (b), denoted as tuples $\left(\bar{g}_{2}^{2},\bar{g}_{3}\right)$,
are given by $\left(0.01,0.1\right)$ (diamonds), $\left(0.09,0.1\right)$
(dots), $\left(0.13,0.066\right)$ (triangles), $\left(0.109,0.04\right)$
(squares). The data shown in panel (c) is obtained by multiplying
the initial values of $\bar{g}_{2}^{2}$ and $\bar{g}_{3}$ by $15$.
In panels (b) and (c) the statistical error in the measurement of
$\bar{g}_{3}\left(\ell\right)$ is represented by error bars. The
same holds for $\bar{g}_{2}\left(\ell\right)$, but the error bars
are too short to be resolved. The variance in panel (d) is computed
for $\left(0.01,0.1\right)$, which is the same as the diamonds in
(b). Again, the error bars are too short to be resolved.

\paragraph*{Figure 2}

The system size for the memory task in panel (a) is $N=32$. Again,
$dt=0.01$ and $T_{\mathrm{therm}}dt=125$. Using a batch size of
$48$ and a learning rate $\eta=0.01$, the training converged after
$5000$ iterations. The process is repeated five times to compute
the average and standard deviation (error bars) of the reconstruction
accuracy as described in \prettyref{subsec:Memory-task}. The stimulus
$s\left(x\right)$ added on top of the thermalized system is a Gaussian
$s\left(x\right)=A\exp\left(-\left(x_{1}+x_{2}\right)^{2}/\left(2\sigma^{2}\right)\right)/\sqrt{2\pi\sigma^{2}}$,
where $A=0.25$ and $\sigma=0.2$. 

For the parity task in panel (b), $T_{\mathrm{therm}}dt=2075$ is
used, and a system of size $N=128$, corresponding to $64$ Fourier
modes in each direction. We use $k_{\mathrm{cut}}=\pi/2$, so that
the lowest $32$ modes are contained in the low-pass filtered snapshots.
Equation \prettyref{eq:response_propagator_momenta_time} shows that
an excited mode decays $\propto\exp\left(-k^{2}t\right)$. The red,
yellow, and green lines in \prettyref{fig:Examples-of-computation.}
(b) show the decay of the modes $k=2\pi n/128,$ where $n\in\left\{ 1,16,32\right\} $,
respectively. The stimuli, encoded in the high modes, introduce fluctuations
on short spatial scales. For the algorithm to remain numerically stable,
it is thus necessary to use a smaller simulation time step $dt$.
Thus, we adapt the time step after the application of the stimulus
from $dt=0.01$ to $dt=0.001$. The values of the interactions are
given by $\bar{g}_{2}=\bar{g}_{3}=0.3$. The signal amplitude is given
by $A=10.0$ and the learning rate $\eta=0.025$. Using a batch size
of $480$, the training converged after $200$ iterations. The mean
and the variance of the share of correctly classified patterns is
computed from eight trials.
\begin{table}
\begin{centering}
\begin{tabular}{ccccccccc}
\hline 
$N$ & $32$ & $48$ & $64$ & $96$ & $128$ & $192$ & $256$ & $384$\tabularnewline
$T_{\mathrm{therm}}dt$ & $125$ & $290$ & $515$ & $1165$ & $2075$ & $4665$ & $8300$ & $2\cdot10^{4}$\tabularnewline
\#samples & $10^{6}$ & $10^{5}$ & $10^{5}$ & $10^{4}$ & $10^{4}$ & $10^{3}$ & $10^{3}$ & $10^{2}$\tabularnewline
\hline 
\end{tabular}
\par\end{centering}
\caption{Summary of the simulation parameters used to produce the results presented
in \prettyref{fig:Phase-diagram} b, c, and d.\label{tab:fig1}}
\end{table}

\section{Computation of the flow equations}

The flow equations \eqref{eq:flow_nu}-\eqref{eq:flow_eta} are computed
to one-loop order within the framework of the infinitesimal momentum
shell Wilsonian RG \citep{Hohenberg77,Medina89,Helias20_445004}.
Once given the Feynman rules Eqs. \eqref{eq:response_propagator},
\eqref{eq:noise_propagator} and \eqref{eq:interaction_vertex}, the
procedure to derive the flow equations is well established. Still,
identifying all the Feynman diagrams contributing to the equations
and their combinatorial factor can be tedious and requires familiarity
with the diagrams' properties that are specific to the model system.
We thus list these diagrams here for completeness. We also calculate
one of the diagrams in detail and use the example to point out properties
of the diagrams that are specific to this field theory.

As an example, we consider the flow of $g_{1}$. This is the coupling
associated with a term $k^{2}\th h$ in the action \prettyref{eq:full_action}.
The one-loop contribution to the flow is thus given by all one-loop
diagrams with exactly one $\th$ and one $h$-leg, that are amputated.
These diagrams are\begin{fmffile}{tpo_app_g1_bare}
\fmfset{thin}{0.75pt}
\fmfset{decor_size}{4mm}
\fmfset{arrow_len}{3mm}
\fmfcmd{style_def wiggly_arrow expr p = cdraw (wiggly p); shrink (0.8); cfill (arrow p); endshrink; enddef;}
\fmfcmd{style_def majorana expr p = cdraw p; cfill (harrow (reverse p, .5)); cfill (harrow (p, .5)) enddef;
		style_def alt_majorana expr p = cdraw p; cfill (tarrow (reverse p, .55)); cfill (tarrow (p, .55)) enddef;}
\begin{align}
	% DIAGRAM 2
	\parbox{30mm}{
		\begin{fmfgraph*}(70,50)
			\fmfleft{o}
			\fmfright{i}
			\fmfcurved
			\fmfsurroundn{v}{2}
			\fmffreeze
			\fmfshift{(0.22w,0.w)}{v2}
			\fmfshift{(-0.22w,0.w)}{v1}
			\fmfdotn{v}{2}
			\fmf{wiggly_arrow, tension=1.0}{v2,o}
			\fmf{wiggly_arrow, tension=1.0}{i,v1}
			\fmf{alt_majorana, left=0.9, tension=0.4, label=$q,,\nu$}{v2,v1}
			\fmf{plain_arrow, left=0.9, tension=0.4, label=$k-q,,\omega - \nu$}{v1,v2}
			\fmflabel{$-k,-\omega$}{o}
			\fmflabel{$k,\omega$}{i}
		\end{fmfgraph*}
	}
	& \hspace{0.7cm}\mathrm{and}\hspace{1.7cm}
	% DIAGRAM 3
	\parbox{30mm}{
		\begin{fmfgraph*}(50,50)
			\fmfstraight
			\fmfbottomn{v}{3}
			\fmffreeze
			\fmfshift{(0.0w,0.3w)}{v1}
			\fmfshift{(0.0w,0.3w)}{v2}
			\fmfshift{(0.0w,0.3w)}{v3}
			\fmfdot{v2}
			\fmf{wiggly_arrow, tension=1.0}{v2,v1}
			\fmf{wiggly_arrow, tension=1.0}{v3,v2}
			\fmf{alt_majorana, right=0.3, tension=0.6, label=$q,,\nu$}{v2,v2}
			\fmflabel{$-k,-\omega$}{v1}
			\fmflabel{$k,\omega$}{v3}
		\end{fmfgraph*}
	}\label{eq:diag_g1_bare}.
\end{align}
\end{fmffile}We are interested in the diagrams' contribution to the term $k^{2}\th h$.
This means we only want to compute them at $0$-th order in $\omega$.
We therefore set $\omega=0$. Also, we want to extract only the $k^{2}$
dependence of the diagrams, neglecting higher orders in $k$. To this
end, we recall that at the critical point $m_{n}=0$. From the vertices
\prettyref{eq:interaction_vertex} we then see that any diagram automatically
carries a $k^{2}$ dependence for each of its external $\th$-legs.
As a result the $k^{2}$ dependence of the diagrams is simply given
by the value of their internal loop at $k=0$.

The diagrams' property above illustrated has two important consequences.
The first is that it considerably simplifies the calculation of diagrams
contributing to the flow of $g_{n}$, as we have seen for $g_{1}$.
The second is that no fluctuation correction contributes to the flow
of couplings associated with terms of $0$-th order in $k$. Indeed,
any term in the action \prettyref{eq:full_action} is at least proportional
to one $\th$ and thus fluctuation corrections are at least of order
$k^{2}$. More specifically, the flow of $D,\,\tau$ and $m_{n}$
is fully determined by mean-field analysis, as stated in the main
text. Notice this is particularly important for the self-consistency
of the whole argument, because we can say with certainty, already
at mean-field level, that $m_{n}=0$ at the critical point.

Coming back to the diagrams \prettyref{eq:diag_g1_bare}, we compute
explicitly the first one. We have to compute it for momenta restricted
to the thin momentum shell $\Lambda/\ell<\left|q\right|<\Lambda$.
This gives\begin{fmffile}{tpo_app_g1_example}
\fmfset{thin}{0.75pt}
\fmfset{decor_size}{4mm}
\fmfset{arrow_len}{3mm}
\fmfcmd{style_def wiggly_arrow expr p = cdraw (wiggly p); shrink (0.8); cfill (arrow p); endshrink; enddef;}
\fmfcmd{style_def majorana expr p = cdraw p; cfill (harrow (reverse p, .5)); cfill (harrow (p, .5)) enddef;
		style_def alt_majorana expr p = cdraw p; cfill (tarrow (reverse p, .55)); cfill (tarrow (p, .55)) enddef;}
\begin{align}
	% DIAGRAM 2
	4\cdot\parbox{26mm}{
		\begin{fmfgraph*}(70,50)
			\fmfleft{o}
			\fmfright{i}
			\fmfcurved
			\fmfsurroundn{v}{2}
			\fmffreeze
			\fmfshift{(0.22w,0.w)}{v2}
			\fmfshift{(-0.22w,0.w)}{v1}
			\fmfdotn{v}{2}
			\fmf{wiggly_arrow, tension=1.0}{v2,o}
			\fmf{wiggly_arrow, tension=1.0}{i,v1}
			\fmf{alt_majorana, left=0.9, tension=0.4, label=$q,,\nu$}{v2,v1}
			\fmf{plain_arrow, left=0.9, tension=0.4, label=$-q,,-\nu$}{v1,v2}
		\end{fmfgraph*}
	\label{eq:diag_g1_example}}
	& =4g_{2}^{2}\int\frac{d^{2}q}{\left(2\pi\right)^{2}}\frac{d\nu}{2\pi}\Delta_{HH}\left(q,\nu\right)\Delta_{H\tilde{H}}\left(q,\nu\right)q^{2}\\
	& =4g_{2}^{2}\int\frac{d^{2}q}{\left(2\pi\right)^{2}}\frac{d\nu}{2\pi}\, \frac{D}{\nu^{2}+g_{1}^2q^{4}}\, \frac{\imath}{\nu-\imath g_{1}q^{2}}\, q^{2}\nonumber\\
	& =-\frac{D}{2\pi}\frac{g_{2}^{2}}{g_{1}}\int_{\Lambda/\ell}^{\Lambda}\frac{dq}{q}=-\frac{D}{2\pi}\frac{g_{2}^{2}}{g_{1}}\ln\ell\,.\nonumber
	% DIAGRAM 3
\end{align}
\end{fmffile}The prefactor in front of the diagram is the combinatorial factor
given by all possible ways of producing the diagram by contracting
the vertices' legs. From the second to the last row, the $\nu$-integral
is computed using the residue theorem of complex analysis. Also the
angular part of the momentum integral is computed, leaving only the
radial part. Finally notice that we make a slight abuse of notation
in \prettyref{eq:diag_g1_example}, as we compute only the internal
loop of the diagram, that is excluding the $k^{2}$ dependence carried
by the external $\th$-leg.

Finally, we are interested in the diagram's contribution to the differential
flow equation \eqref{eq:flow_nu}, that is to $\frac{1}{g_{1}}\frac{dg_{1}}{ds}$,
with $s\coloneqq\left(2\pi\right)^{-1}\ln\left(\ell\right)$. Thus
we compute\begin{fmffile}{tpo_app_g1_example_cont}
\fmfset{thin}{0.75pt}
\fmfset{decor_size}{4mm}
\fmfset{arrow_len}{3mm}
\fmfcmd{style_def wiggly_arrow expr p = cdraw (wiggly p); shrink (0.8); cfill (arrow p); endshrink; enddef;}
\fmfcmd{style_def majorana expr p = cdraw p; cfill (harrow (reverse p, .5)); cfill (harrow (p, .5)) enddef;
		style_def alt_majorana expr p = cdraw p; cfill (tarrow (reverse p, .55)); cfill (tarrow (p, .55)) enddef;}
\begin{align}
	% DIAGRAM 2
	\frac{2\pi}{g_{1}}\ell\frac{d}{d\ell}4\cdot\parbox{26mm}{
		\begin{fmfgraph*}(70,50)
			\fmfleft{o}
			\fmfright{i}
			\fmfcurved
			\fmfsurroundn{v}{2}
			\fmffreeze
			\fmfshift{(0.22w,0.w)}{v2}
			\fmfshift{(-0.22w,0.w)}{v1}
			\fmfdotn{v}{2}
			\fmf{wiggly_arrow, tension=1.0}{v2,o}
			\fmf{wiggly_arrow, tension=1.0}{i,v1}
			\fmf{alt_majorana, left=0.9, tension=0.4}{v2,v1}
			\fmf{plain_arrow, left=0.9, tension=0.4}{v1,v2}
		\end{fmfgraph*}
	\label{eq:diag_g1_example_cont}}
	& =-\frac{g_{2}^{2}}{g_{1}^{2}}\frac{D}{g_{1}}=-\bar{g}_{2}^{2}\,,
	% DIAGRAM 3
\end{align}
\end{fmffile}which is indeed the second term on the r.h.s. of \prettyref{eq:flow_nu}.
We left implicit that the derivative by $\ell$ in \prettyref{eq:diag_g1_example_cont}
must be computed at $\ell=1$, which is the limit of an infinitesimal
momentum shell.

In the following we give all the other diagrams contributing to the
flow equations. There are two diagrams contributing to the flow of
$g_{1}$:\begin{fmffile}{tpo_app_g1}
\fmfset{thin}{0.75pt}
\fmfset{decor_size}{4mm}
\fmfset{arrow_len}{3mm}
\fmfcmd{style_def wiggly_arrow expr p = cdraw (wiggly p); shrink (0.8); cfill (arrow p); endshrink; enddef;}
\fmfcmd{style_def majorana expr p = cdraw p; cfill (harrow (reverse p, .5)); cfill (harrow (p, .5)) enddef;
		style_def alt_majorana expr p = cdraw p; cfill (tarrow (reverse p, .55)); cfill (tarrow (p, .55)) enddef;}
\begin{align*}
	% DIAGRAM 2
	\ell \frac{d}{d\ell} \cdot 4 \cdot \,\parbox{26mm}{
		\begin{fmfgraph*}(70,50)
			\fmfleft{o}
			\fmfright{i}
			\fmfcurved
			\fmfsurroundn{v}{2}
			\fmffreeze
			\fmfshift{(0.22w,0.w)}{v2}
			\fmfshift{(-0.22w,0.w)}{v1}
			\fmfdotn{v}{2}
			\fmf{wiggly_arrow, tension=1.0}{v2,o}
			\fmf{wiggly_arrow, tension=1.0}{i,v1}
			\fmf{alt_majorana, left=0.9, tension=0.4}{v2,v1}
			\fmf{plain_arrow, left=0.9, tension=0.4}{v1,v2}
		\end{fmfgraph*}
	}
	&= -\frac{D}{2\pi} \frac{g_2^2}{g_1^2} & \mathrm{and} &&
	% DIAGRAM 3
	\ell \frac{d}{d\ell} \cdot 3 \cdot \,\parbox{20mm}{
		\begin{fmfgraph*}(50,50)
			\fmfstraight
			\fmfbottomn{v}{3}
			\fmffreeze
			\fmfshift{(0.0w,0.3w)}{v1}
			\fmfshift{(0.0w,0.3w)}{v2}
			\fmfshift{(0.0w,0.3w)}{v3}
			\fmfdot{v2}
			\fmf{wiggly_arrow, tension=1.0}{v2,v1}
			\fmf{wiggly_arrow, tension=1.0}{v3,v2}
			\fmf{alt_majorana, right=0.3, tension=0.6}{v2,v2}
		\end{fmfgraph*}
	}
	&= \frac{3}{2} \frac{D}{2\pi} \frac{g_3}{g_1}.
	\\
\end{align*}
\end{fmffile}The following four diagrams contribute to the flow of $g_{2}$:\begin{fmffile}{tpo_app_g2}
\fmfset{thin}{0.75pt}
\fmfset{decor_size}{4mm}
\fmfset{arrow_len}{3mm}
\fmfcmd{style_def wiggly_arrow expr p = cdraw (wiggly p); shrink (0.8); cfill (arrow p); endshrink; enddef;}
\fmfcmd{style_def majorana expr p = cdraw p; cfill (harrow (reverse p, .5)); cfill (harrow (p, .5)) enddef;
		style_def alt_majorana expr p = cdraw p; cfill (tarrow (reverse p, .55)); cfill (tarrow (p, .55)) enddef;}
\begin{align*}
		% DIAGRAM 4
	\ell \frac{d}{d\ell} \cdot 4 \cdot \,\parbox{22mm}{
	\begin{fmfgraph*}(60,60)
			\fmftopn{i}{2}
			\fmfbottom{o}
			\fmfcurved
			\fmfsurroundn{v}{3}
			\fmffreeze
			\fmfshift{(-0.2w,0.2w)}{v1}
			\fmfshift{(-0.0w,-0.25w)}{v2}
			\fmfshift{(0.265w,0.2w)}{v3}
			\fmfdotn{v}{3}
			\fmf{wiggly_arrow, tension=1.0}{v3,o}
			\fmf{wiggly_arrow, tension=1.0}{i1,v2}
			\fmf{wiggly_arrow, tension=1.0}{i2,v1}
			\fmf{plain_arrow, left=0.5, tension=0.4}{v1,v3}
			\fmf{plain_arrow, right=0.5, tension=0.4}{v2,v3}
			\fmf{alt_majorana, left=0.5, tension=0.4}{v2,v1}
		\end{fmfgraph*}
	}
	&= -\frac{D}{2\pi} \frac{g_2^3}{g_1^3}\, , && &
	% DIAGRAM 5
	\ell \frac{d}{d\ell} \cdot 8 \cdot \,\parbox{22mm}{
		\begin{fmfgraph*}(60,60)
			\fmftopn{i}{2}
			\fmfbottom{o}
			\fmfcurved
			\fmfsurroundn{v}{3}
			\fmffreeze
			\fmfshift{(-0.2w,0.2w)}{v1}
			\fmfshift{(-0.0w,-0.25w)}{v2}
			\fmfshift{(0.265w,0.2w)}{v3}
			\fmfdotn{v}{3}
			\fmf{wiggly_arrow, tension=1.0}{v3,o}
			\fmf{wiggly_arrow, tension=1.0}{i1,v2}
			\fmf{wiggly_arrow, tension=1.0}{i2,v1}
			\fmf{plain_arrow, left=0.5, tension=0.4}{v2,v1,v3}
			\fmf{alt_majorana, left=0.5, tension=0.4}{v3,v2}
		\end{fmfgraph*}
	}
	&= -\frac{D}{2\pi} \frac{g_2^3}{g_1^3}\, ,
	\\
	% DIAGRAM 6
	\ell \frac{d}{d\ell} \cdot 8 \cdot \,\parbox{22mm}{
		\begin{fmfgraph*}(60,60)
			\fmftopn{i}{2}
			\fmfbottom{o}
			\fmfleftn{v}{2}
			\fmffreeze
			\fmfshift{(0.4w,0.25w)}{v1}
			\fmfshift{(0.4w,-0.2w)}{v2}
			\fmfdotn{v}{2}
			\fmf{wiggly_arrow, tension=1.0}{v1,o}
			\fmf{wiggly_arrow, tension=1.0}{i1,v2}
			\fmf{wiggly_arrow, tension=1.0}{i2,v2}
			\fmf{plain_arrow, left=0.5, tension=0.4}{v2,v1}
			\fmf{alt_majorana, left=0.5, tension=0.4}{v1,v2}
		\end{fmfgraph*}
	}
	&= -\frac{D}{2\pi} \frac{g_2 g_3}{g_1^2}\, , &\mathrm{and}&&
	% DIAGRAM 7
	\ell \frac{d}{d\ell} \cdot 12 \cdot \,\parbox{22mm}{
		\begin{fmfgraph*}(60,60)
			\fmftopn{i}{2}
			\fmfbottom{o}
			\fmfleftn{v}{2}
			\fmffreeze
			\fmfshift{(0.4w,0.25w)}{v1}
			\fmfshift{(0.4w,-0.2w)}{v2}
			\fmfshift{(-0.1w,-0.4w)}{i2}
			\fmfdotn{v}{2}
			\fmf{wiggly_arrow, tension=1.0}{v1,o}
			\fmf{wiggly_arrow, tension=1.0}{i1,v2}
			\fmf{wiggly_arrow, tension=1.0}{i2,v1}
			\fmf{plain_arrow, left=0.5, tension=0.4}{v2,v1}
			\fmf{alt_majorana, left=0.5, tension=0.4}{v1,v2}
		\end{fmfgraph*}
	}
	&= -3\frac{D}{2\pi} \frac{g_2 g_3}{g_1^2}.
	\\
\end{align*}
\end{fmffile}Finally, eight diagrams contribute to the flow of $g_{3}$:% four-point interaction vertex
\begin{fmffile}{tpo_app_g3}
\fmfset{thin}{0.75pt}
\fmfset{decor_size}{4mm}
\fmfset{arrow_len}{3mm}
\fmfcmd{style_def wiggly_arrow expr p = cdraw (wiggly p); shrink (0.8); cfill (arrow p); endshrink; enddef;}
\fmfcmd{style_def majorana expr p = cdraw p; cfill (harrow (reverse p, .5)); cfill (harrow (p, .5)) enddef;
		style_def alt_majorana expr p = cdraw p; cfill (tarrow (reverse p, .55)); cfill (tarrow (p, .55)) enddef;}
\begin{align*}
		% DIAGRAM 8
	\ell \frac{d}{d\ell} \cdot 16 \cdot \,\parbox{22mm}{
	\begin{fmfgraph*}(60,60)
			\fmfcurved
			\fmfsurroundn{o}{4}
			\fmfsurroundn{i}{4}
			\fmffreeze
			\fmfshift{(0.0w,0.25w)}{i4}
			\fmfshift{(-0.25w,0.0w)}{i1}
			\fmfshift{(0.0w,-0.25w)}{i2}
			\fmfshift{(0.25w,0.0w)}{i3}
			\fmfdotn{i}{4}
			\fmf{wiggly_arrow, tension=1.0}{i4,o4}
			\fmf{wiggly_arrow, tension=1.0}{o1,i1}
			\fmf{wiggly_arrow, tension=1.0}{o3,i3}
			\fmf{wiggly_arrow, tension=1.0}{o2,i2}
			\fmf{plain_arrow, right=0.5, tension=0.4}{i1,i2,i3,i4}
			\fmf{alt_majorana, right=0.5, tension=0.4}{i4,i1}
		\end{fmfgraph*}
	}
	&= -\frac{D}{2\pi} \frac{g_2^4}{g_1^4}\, , && &
	% DIAGRAM 9
	\ell \frac{d}{d\ell} \cdot 16 \cdot \,\parbox{22mm}{
	\begin{fmfgraph*}(60,60)
			\fmfcurved
			\fmfsurroundn{o}{4}
			\fmfsurroundn{i}{4}
			\fmffreeze
			\fmfshift{(0.0w,0.25w)}{i4}
			\fmfshift{(-0.25w,0.0w)}{i1}
			\fmfshift{(0.0w,-0.25w)}{i2}
			\fmfshift{(0.25w,0.0w)}{i3}
			\fmfdotn{i}{4}
			\fmf{wiggly_arrow, tension=1.0}{i4,o4}
			\fmf{wiggly_arrow, tension=1.0}{o1,i1}
			\fmf{wiggly_arrow, tension=1.0}{o3,i3}
			\fmf{wiggly_arrow, tension=1.0}{o2,i2}
			\fmf{plain_arrow, right=0.5}{i2,i3,i4}
			\fmf{alt_majorana, right=0.5}{i1,i2}
			\fmf{plain_arrow, left=0.5}{i1,i4}
		\end{fmfgraph*}
	}
	&= -3 \frac{D}{2\pi} \frac{g_2^4}{g_1^4}\, ,
	\\
	% DIAGRAM 10
	\ell \frac{d}{d\ell} \cdot 12 \cdot \,\parbox{22mm}{
		\begin{fmfgraph*}(60,60)
			\fmftopn{i}{2}
			\fmfbottom{o}
			\fmfright{i3}
			\fmfcurved
			\fmfsurroundn{v}{3}
			\fmffreeze
			\fmfshift{(-0.2w,0.2w)}{v1}
			\fmfshift{(-0.0w,-0.25w)}{v2}
			\fmfshift{(0.265w,0.2w)}{v3}
			\fmfshift{(0.0w,-0.2w)}{i3}
			\fmfdotn{v}{3}
			\fmf{wiggly_arrow, tension=1.0}{v3,o}
			\fmf{wiggly_arrow, tension=1.0}{i1,v2}
			\fmf{wiggly_arrow, tension=1.0}{i2,v1}
			\fmf{wiggly_arrow, tension=1.0}{i3,v3}
			\fmf{plain_arrow, left=0.5, tension=0.4}{v1,v3}
			\fmf{plain_arrow, right=0.5, tension=0.4}{v2,v3}
			\fmf{alt_majorana, right=0.5, tension=0.4}{v1,v2}
		\end{fmfgraph*}
	}
	&= 3\frac{D}{2\pi} \frac{g_2^2 g_3}{g_1^3}\, , && &
	% DIAGRAM 15
	\ell \frac{d}{d\ell} \cdot 24 \cdot \,\parbox{22mm}{
		\begin{fmfgraph*}(60,60)
			\fmftopn{i}{2}
			\fmfbottom{o}
			\fmfright{i3}
			\fmfcurved
			\fmfsurroundn{v}{3}
			\fmffreeze
			\fmfshift{(-0.2w,0.2w)}{v1}
			\fmfshift{(-0.0w,-0.25w)}{v2}
			\fmfshift{(0.265w,0.2w)}{v3}
			\fmfshift{(0.0w,-0.2w)}{i3}
			\fmfdotn{v}{3}
			\fmf{wiggly_arrow, tension=1.0}{v3,o}
			\fmf{wiggly_arrow, tension=1.0}{i1,v2}
			\fmf{wiggly_arrow, tension=1.0}{i2,v1}
			\fmf{wiggly_arrow, tension=1.0}{i3,v3}
			\fmf{plain_arrow, left=0.5, tension=0.4}{v2,v1,v3}
			\fmf{alt_majorana, left=0.5, tension=0.4}{v3,v2}
		\end{fmfgraph*}
	}
	&= 3 \frac{D}{2\pi} \frac{g_2^2 g_3}{g_1^3}\, ,
	\\
	% DIAGRAM 11
	\ell \frac{d}{d\ell} \cdot 12 \cdot \,\parbox{22mm}{
		\begin{fmfgraph*}(60,60)
			\fmftopn{i}{2}
			\fmfbottom{o}
			\fmftop{i3}
			\fmfcurved
			\fmfsurroundn{v}{3}
			\fmffreeze
			\fmfshift{(-0.2w,0.2w)}{v1}
			\fmfshift{(-0.0w,-0.25w)}{v2}
			\fmfshift{(0.265w,0.2w)}{v3}
			\fmfshift{(0.2w,0.0w)}{i3}
			\fmfshift{(0.0w,-0.1w)}{i2}
			\fmfdotn{v}{3}
			\fmf{wiggly_arrow, tension=1.0}{v3,o}
			\fmf{wiggly_arrow, tension=1.0}{i1,v2}
			\fmf{wiggly_arrow, tension=1.0}{i2,v1}
			\fmf{wiggly_arrow, tension=1.0}{i3,v1}
			\fmf{plain_arrow, left=0.5, tension=0.4}{v1,v3}
			\fmf{plain_arrow, right=0.5, tension=0.4}{v2,v3}
			\fmf{alt_majorana, right=0.5, tension=0.4}{v1,v2}
		\end{fmfgraph*}
	}
	&= 3\frac{D}{2\pi} \frac{g_2^2 g_3}{g_1^3}\, , && &
	% DIAGRAM 12
	\ell \frac{d}{d\ell} \cdot 12 \cdot \,\parbox{22mm}{
		\begin{fmfgraph*}(60,60)
			\fmftopn{i}{2}
			\fmfbottom{o}
			\fmftop{i3}
			\fmfcurved
			\fmfsurroundn{v}{3}
			\fmffreeze
			\fmfshift{(-0.2w,0.2w)}{v1}
			\fmfshift{(-0.0w,-0.25w)}{v2}
			\fmfshift{(0.265w,0.2w)}{v3}
			\fmfshift{(0.2w,0.0w)}{i3}
			\fmfshift{(0.0w,-0.1w)}{i2}
			\fmfdotn{v}{3}
			\fmf{wiggly_arrow, tension=1.0}{v3,o}
			\fmf{wiggly_arrow, tension=1.0}{i1,v2}
			\fmf{wiggly_arrow, tension=1.0}{i2,v1}
			\fmf{wiggly_arrow, tension=1.0}{i3,v1}
			\fmf{plain_arrow, right=0.5, tension=0.4}{v1,v2,v3}
			\fmf{alt_majorana, right=0.5, tension=0.4}{v3,v1}
		\end{fmfgraph*}
	}
	&= \frac{3}{2} \frac{D}{2\pi} \frac{g_2^2 g_3}{g_1^3}\, ,
	\\
% DIAGRAM 13
	\ell \frac{d}{d\ell} \cdot 12 \cdot \,\parbox{22mm}{
		\begin{fmfgraph*}(60,60)
			\fmftopn{i}{2}
			\fmfbottom{o}
			\fmftop{i3}
			\fmfcurved
			\fmfsurroundn{v}{3}
			\fmffreeze
			\fmfshift{(-0.2w,0.2w)}{v1}
			\fmfshift{(-0.0w,-0.25w)}{v2}
			\fmfshift{(0.265w,0.2w)}{v3}
			\fmfshift{(0.2w,0.0w)}{i3}
			\fmfshift{(0.0w,-0.1w)}{i2}
			\fmfdotn{v}{3}
			\fmf{wiggly_arrow, tension=1.0}{v3,o}
			\fmf{wiggly_arrow, tension=1.0}{i1,v2}
			\fmf{wiggly_arrow, tension=1.0}{i2,v1}
			\fmf{wiggly_arrow, tension=1.0}{i3,v1}
			\fmf{plain_arrow, left=0.5, tension=0.4}{v2,v1,v3}
			\fmf{alt_majorana, left=0.5, tension=0.4}{v3,v2}
		\end{fmfgraph*}
	}
	&= \frac{3}{2} \frac{D}{2\pi} \frac{g_2^2 g_3}{g_1^3}\, , &\mathrm{and}&&
	% DIAGRAM 14
	\ell \frac{d}{d\ell} \cdot 12 \cdot \,\parbox{22mm}{
		\begin{fmfgraph*}(60,60)
			\fmftopn{i}{2}
			\fmfright{i3}
			\fmfbottom{o}
			\fmfleftn{v}{2}
			\fmffreeze
			\fmfshift{(0.4w,0.25w)}{v1}
			\fmfshift{(0.4w,-0.2w)}{v2}
			\fmfshift{(0.0w,-0.2w)}{i3}
			\fmfdotn{v}{2}
			\fmf{wiggly_arrow, tension=1.0}{v1,o}
			\fmf{wiggly_arrow, tension=1.0}{i1,v2}
			\fmf{wiggly_arrow, tension=1.0}{i3,v1}
			\fmf{wiggly_arrow, tension=1.0}{i2,v2}
			\fmf{plain_arrow, left=0.5, tension=0.4}{v2,v1}
			\fmf{alt_majorana, left=0.5, tension=0.4}{v1,v2}
		\end{fmfgraph*}
	}
	&= -\frac{9}{2}\frac{D}{2\pi} \frac{g_2 g_3}{g_1^2}.
	\\
\end{align*}
\end{fmffile}

\end{appendices}

\bibliographystyle{apsrev_brain}
\bibliography{brain}

\makeatletter\@input{xpaper.tex}\makeatother